 \documentstyle[preprint,aps,eqsecnum]{revtex}
 \tighten
\begin{document}
\draft
\hyphenation{Rijken}
\hyphenation{Nijmegen}

\title{Soft two-meson-exchange nucleon-nucleon potentials.
       \\ I. Planar and crossed-box diagrams}

\author{Th.A.\ Rijken}
\address{Institute for Theoretical Physics, University of Nijmegen,
         Nijmegen, The Netherlands}

\author{V.G.J.\ Stoks\footnote{Present address:
        TRIUMF, 4004 Wesbrook Mall, Vancouver, British Columbia,
        Canada V6T 2A3}}
\address{Department of Physics,
         The Flinders University of South Australia,
         Bedford Park, South Australia 5042, Australia}

\date{version of: \today}
\maketitle

\begin{abstract}
Pion-meson-exchange nucleon-nucleon potentials are derived
for two nucleons in the intermediate states. The mesons we include
are (i) pseudoscalar mesons: $\pi, \eta, \eta'$;
(ii) vector mesons: $\rho, \omega, \phi$;
(iii) scalar mesons: $a_{0}(980), \varepsilon(760), f_{0}(975)$;
and (iv) the $J=0$ contribution from the Pomeron.
Strong dynamical pair suppression is assumed, and at the
nucleon-nucleon-meson vertices Gaussian form factors are incorporated
into the relativistic two-body framework using a dispersion
representation for the pion- and meson-exchange amplitudes.
The Fourier transformations are performed using factorization
techniques for the energy denominators. The potentials are first
calculated in the adiabatic approximation to all planar and crossed
three-dimensional momentum-space $\pi$-meson diagrams. Next, we
calculate the $1/M$ corrections.
\end{abstract}

\narrowtext

\section{INTRODUCTION}
\label{sec:chap1}
This paper and its companion paper~\cite{Rij95b} are part of our
program to extend the Nijmegen soft-core one-boson-exchange
potential~\cite{Nag78} by including two-meson-exchange potentials.
The material presented here is incorporated into a new extended
soft-core nucleon-nucleon model, hereafter referred to as the ESC
potential. We are still investigating ways to improve this model as
much as possible, but the preliminary version of the ESC potential
employed in the present paper is here used to illustrate the
contributions of the various two-meson exchanges.

In low-energy nucleon-nucleon models, exchanges with an effective
mass less than 1 GeV seem to be the most important ones. The very
heavy mesons possibly only play some role for $S$ waves, but they can
presumably be covered up by the lower-mass mesons. After one-meson
exchange, the effects of two-meson exchanges at low energies seem
worthwhile to be investigated and, as we will demonstrate with the
ESC potential, they indeed provide a major improvement in the
theoretical description of the nucleon-nucleon data.

Recently, we have evaluated the soft two-pion-exchange potentials in
coordinate space, using some new techniques~\cite{Rij91}. We noted
that these techniques could also be applied to double-meson exchanges
where the mesons have different masses.
For the double-meson exchanges we can distinguish two different
classes. The first class consists of the diagrams with two baryons
(nucleons) in the intermediate states, which can be simply understood
as the second-order contributions in a series expansion of multi-meson
exchanges. In this paper we calculate these two-meson-exchange (TME)
potentials, using Gaussian form factors, for the cases where one of
the exchanged mesons is the pion. For exchanges with effective masses
below 1 GeV these are the most important ones. An extension to the
most general two-meson exchange (not necessarily involving at least
one pion), can then be evaluated in a straight forward manner.
Furthermore, we restrict ourselves to the case with two nucleons in
the intermediate states. The potentials containing the isobar
excitation effects explicitly have been considered
elsewhere~\cite{Rij92a,Rij92b}.

The second class consists of the diagrams where either one or both
nucleons contains a pair vertex (``seagull'' diagrams).
The pair-meson diagrams can be viewed as the result of
integrating out the heavy-meson and resonance degrees of freedom
in the two-meson-exchange processes (see, e.g., Ref.~\cite{Wei90}).
It will appear that these ``seagull'' contributions can be interpreted
successfully in a ``dual'' picture~\cite{Dol68}. The various
possibilities for meson pairs coupling to the nucleon are inspired
by the chiral-invariant phenomenological Lagrangians that have
appeared in the literature.
The pair-meson potentials belonging to this second class will be
presented in the companion paper~\cite{Rij95b}.

The mesons we include are (i) pseudoscalar mesons: $\pi$, $\eta, \eta'$;
(ii) vector mesons: $\rho, \omega, \phi$; (ii) scalar mesons~\cite{eps}:
$a_{0}(980), \varepsilon(760), f_{0}(975)$; and (iv) the $J=0$
contribution from the Pomeron. The $2\pi$-exchange potential has
already been derived in Ref.~\cite{Rij91}, and we here only include
those results for reasons of completeness.
In the literature, the treatment of two-meson exchanges is incomplete.
For example, the $\pi\omega$ potential has been derived in the
dispersion-relation approach~\cite{Ris76}, which does not allow for
the inclusion of Gaussian form factors as applied in the present
study. For the $\pi\omega$, $\pi\rho$, and $\pi\sigma$ exchanges, only
the non-iterative potentials from the graphs with at least one meson in
the intermediate state (see below) have been evaluated in momentum space
in the framework of noncovariant perturbation theory~\cite{Hol81}.
In this paper we give, apart from the $2\pi$-exchange, the potentials
for all pion-meson exchanges with a total mass less than $\sim 1$ GeV.

Our general approach to two-meson-exchange potentials is given, in
principle, in Ref.~\cite{Rij91}. Starting from the relativistic two-body
equations, the two-meson-exchange potentials for the relativistic
three-dimensional integral equation and for the Lippmann-Schwinger
equation are derived from the second-order Feynman diagrams by
integration over the relative energies~\cite{Sal52}. This way, the
``old-fashioned'' perturbation graphs for the TME potentials are
obtained from the Feynman diagrams.

The proper way to incorporate Gaussian form factors is explained in
\cite{Rij91} and, as already mentioned above, in this paper the
important ``factorization'' technique involving two mesons with
different masses will be exploited. It is in particular this
factorization technique which enables us to express the two-meson
exchanges in terms of the basic coordinate-space function
$\phi^{0}_{C}(r)$ of the Nijmegen one-boson-exchange (OBE)
model~\cite{Nag78,Mae89,Sto94}. Technically, there is no compelling
reason to work in coordinate space, rather than in momentum space.
Electromagnetic effects can be included more easily in coordinate
space, whereas relativistic effects can be studied more effectively
in momentum space. However, we think that at low energies relativistic
effects can be included sufficiently accurate in coordinate space.
Coordinate-space interactions are useful, for example, in Jastrow
variational calculations (see, e.g., Refs.~\cite{Wir91,Pie92}) or in
applications where one wishes to distinguish a (long-range)
meson-exchange region and a (short-range) quark-exchange region
(see, e.g., Refs.~\cite{Tak89,Fer93}). Furthermore, the effects of
the particular shape of the central, spin-spin, tensor, and spin-orbit
components on the phase shifts is much more transparent in a
coordinate-space picture.
Ultimately, it is our goal to produce an $N\!N$ potential which is
exactly equivalent in both coordinate space and momentum space.

The graphs which we calculate are: (i) the planar and crossed
$\pi$-meson-exchange graphs of the similar type as those calculated
by Brueckner and Watson~\cite{Bru53} for two-pion exchange, and
(ii) the iterated $\pi$-meson-exchange graphs of the type of graphs
calculated by Taketani, Machida, and Ohnuma~\cite{Tak52} for two-pion
exchange.
As the distinction between these two classes of graphs is convenient
as a means to denote the different contributions, we will also adopt
in this paper the following nomenclature: the non-iterative potential
graphs with at least one meson in the intermediate state are of class
(i) and will be referred to as planar ($//$) and crossed ($X$) BW
graphs, while the iterative graphs with only nucleons in the intermediate
state are of class (ii) and will be referred to as TMO graphs.

The $\pi$-meson-exchange potentials are calculated first using the
adiabatic approximation for the nucleons in the intermediate states.
However, in the meson exchanges we include the $1/M^{2}$ terms, in
analogy with the OBE model~\cite{Nag78,Mae89,Sto94}. We find that up
to this order the $\pi\varepsilon$ and $\pi f_{0}$ potentials are
identically zero. Also, in the $\pi$-vector and $\pi$-pseudoscalar
potentials there are large cancelations between the contributions from
the planar and crossed-box diagrams when the meson is an isoscalar
($I=0$) meson.
Therefore, the next to leading order is expected to be important, and
so we here also calculate the nonadiabatic $1/M$ corrections and the
$1/M$ corrections from the pseudovector coupling of the pion and the
other pseudoscalar mesons.

The paper is organized as follows.
In Sec.~\ref{sec:chap2} we define the two-meson-exchange kernels.
We give the interaction Hamiltonians and briefly indicate how the
Gaussian form factors can be implemented. In Sec.~\ref{sec:chap3}
the definition of the nucleon-nucleon two-meson potential for the
Lippmann-Schwinger equation is briefly repeated from \cite{Rij91}
and the vertices in Pauli-spinor space for the different couplings are
given. Here we also mention the approximations made in these vertices.
In the next section, the $\pi$-pseudoscalar, the $\pi$-scalar,
the $\pi$-Pomeron, and the $\pi$-vector potentials are derived.
In Sec.~\ref{sec:chap5} we derive the $1/M$ corrections, due to (i)
the nonadiabatic corrections, (ii) the $1/M$ term in the pseudovector
coupling of the pion and the other pseudoscalar mesons, and (iii)
a cancelation between the $1/M^{2}$ off-shell term in the pseudovector
vertices and the $1/M$ term of the two-nucleon intermediate state
in the TMO diagrams.
In Sec.~\ref{sec:chap7} the results are shown and discussed.

Appendix~\ref{app:appA} contains a dictionary of differentiation
formulas, adequate for deriving the final form of the potentials of
this paper in coordinate space. Explicit expressions for the
leading-order contributions to the coordinate-space potentials are
given in Appendix~\ref{app:appB}.

\section{THE TWO-MESON-EXCHANGE KERNEL}
\label{sec:chap2}
The fourth-order two-meson-exchange kernel is derived following
the procedure as discussed in \cite{Rij92b}, to which we
refer for details and definitions. The only difference is that here
we choose to shift the $M/E$ factors into the Green's function.
This means that the Thompson equation reads
\begin{eqnarray}
   \phi_{++}({\bf p}')=&&\phi^{(0)}_{++}({\bf p}') +
        E_{2}^{(+)}({\bf p}';W)  \nonumber\\
   &&\times \int\!d^{3}p K^{\rm irr}({\bf p}',{\bf p}|W)
        \phi_{++}({\bf p}),         \label{Thompson}
\end{eqnarray}
where now the Green's function is
\begin{equation}
   E_{2}^{(+)}({\bf p}';W)=\frac{1}{(2\pi)^{3}}
        \left[\frac{M_{a}M_{b}}{E_{a}({\bf p}')E_{b}({\bf p}')}\right]
        \frac{\Lambda_{+}^{a}({\bf p}')\Lambda_{+}^{b}(-{\bf p}')}
        {W-{\cal W}({\bf p}')+i\delta},   \label{Ethompson}
\end{equation}
and so the irreducible kernel is then given by
\widetext
\begin{eqnarray}
  K^{\rm irr}({\bf p}',{\bf p}|W)=-&&(2\pi)^{-2}
    [W-{\cal W}({\bf p}')]\;[W-{\cal W}({\bf p})]
    \int_{-\infty}^{\infty}\!dp_{0}'\int_{-\infty}^{\infty}\!dp_{0}
                                                 \nonumber\\
    &&\times\, \left\{\left[F_{W}^{(a)}({\bf p}',p_{0}')
      F_{W}^{(b)}(-{\bf p}',-p_{0}')\right]^{-1}
      \left[I(p_{0}',{\bf p}';p_{0},{\bf p})\right]_{++,++}
\right.  \nonumber\\ &&\times \left.
      \left[F_{W}^{(a)}({\bf p},p_{0})F_{W}^{(b)}(-{\bf p},-p_{0})
      \right]^{-1}\right\}.
\end{eqnarray}
\narrowtext \noindent
The transition from the Thompson equation (\ref{Thompson}) to the
Lippmann-Schwinger equation,
\begin{equation}
   \phi({\bf p}')=\phi^{(0)}({\bf p}') + g({\bf p}';W)
      \int\!d^{3}p V({\bf p}',{\bf p}|W) \phi({\bf p}), \label{LSeq}
\end{equation}
is effectuated by the transformations
\begin{eqnarray}
   &&\phi_{++}({\bf p})=N({\bf p};W)\phi({\bf p}),   \nonumber\\
   &&K^{\rm irr}({\bf p}',{\bf p}|W)=N^{-1}({\bf p}';W)
     V({\bf p}',{\bf p}|W)N^{-1}({\bf p};W),         \nonumber\\
   &&E_{2}^{(+)}({\bf p};W)=N^{2}({\bf p};W)g({\bf p};W), \label{trans}
\end{eqnarray}
with the nonrelativistic Green's function
\begin{equation}
  g({\bf p};W)=\frac{1}{(2\pi)^{3}}\Lambda_{+}^{a}({\bf p})
     \Lambda_{+}^{b}(-{\bf p})
     \frac{M}{{\bf p}_{i}^{2}-{\bf p}^{2}+i\delta},
\end{equation}
where ${\bf p}_{i}$ is the on-shell momentum.
For $N\!N$ scattering the equal-mass approximation is very nearly valid.
In that case we have, taking the nucleons on the energy shell,
\begin{equation}
   N({\bf p};W) = \sqrt{M/E({\bf p})}.      \label{Nfactor}
\end{equation}
Taking further into account the normalization factors of the nucleon
Dirac spinors, it is then natural to write
\begin{eqnarray}
   V({\bf p}',{\bf p})&=&\sqrt{\frac{M}{E({\bf p}')}}
     \left(\frac{E({\bf p}')+M}{2M}\right)U({\bf p}',{\bf p}) \nonumber\\
   && \times \left(\frac{E({\bf p})+M}{2M}\right)
     \sqrt{\frac{M}{E({\bf p})}}     \nonumber\\
   &\approx& U({\bf p}',{\bf p})[1 + {\cal O}(1/M^{4})],    \label{Upot}
\end{eqnarray}
where the potential function $U({\bf p}',{\bf p})$ has a reduced
energy dependence.

The two-meson-exchange kernel is written as a power series in $\lambda$,
which denotes the number of nucleon-nucleon-meson ($N\!Nm$) vertices.
The various two-meson contributions to the fourth-order $\lambda$
terms of the two-meson-exchange (TME) kernel define the various
two-meson-exchange potentials. The corresponding fourth-order elastic
$N\!N$ matrix element of the kernel now reads
\widetext
\begin{eqnarray}
  K^{(4)}({\bf p}',{\bf p}|W)_{a'b';ab}&=&-(2\pi)^{-2}
         [W-{\cal W}({\bf p}')]\;[W-{\cal W}({\bf p})]
\nonumber\\ &\times&
    \sum_{a'',b''}\int\!dp_{0}'\int\!dp_{0}\int\!dk_{10}\int\!dk_{20}
                  \int\!d{\bf k}_{1}\int\!d{\bf k}_{2}      %\nonumber\\
                \ i(2\pi)^{-4}\delta^{4}(p'-p-k_{1}-k_{2})
\nonumber\\ &\times&
   [k^{2}_{2}-m^{2}_{2}+i\delta]^{-1}\left[F^{(a')}_{W}({\bf p}',p_{0}')
       F^{(b')}_{W}(-{\bf p}',-p_{0}') \right]^{-1}          \nonumber\\
    &\times& \left\{ [\Gamma_{j} F_{W}^{-1}
          ({\bf p}+{\bf k}_{1}, p_{0}+k_{10}) \Gamma_{i}]^{(a'')}
          [\Gamma_{j} F_{W}^{-1}(-{\bf p}-{\bf k}_{1},-p_{0}-k_{10})
           \Gamma_{i}]^{(b'')}    \right.                    \nonumber\\
    && + \left. [\Gamma_{j} F_{W}^{-1}
          ({\bf p}+{\bf k}_{1}, p_{0}+k_{10}) \Gamma_{i}]^{(a'')}
          [\Gamma_{i}\ F_{W}^{-1}(-{\bf p}'+{\bf k}_{1},
          -p'_{0}+k_{10})\ \Gamma_{j}]^{(b'')} \right\}     \nonumber\\
    &\times& \left[F^{(a)}_{W}({\bf p},p_{0})F^{(b)}_{W}
          (-{\bf p},-p_{0})\right]^{-1}[k^{2}_{1}-m^{2}_{1}+i\delta]^{-1}.
                                          \label{ker4eq}
\end{eqnarray}
\narrowtext \noindent
Here $m_{1}$ and $m_{2}$ denote the two meson masses,
and the $\Gamma_{i}$ and $\Gamma_{j}$ denote the nucleon-nucleon-meson
vertices, which follow from the interaction Hamiltonians (see below).
Because here we only consider nucleons in the intermediate state,
we have $a=a'=a''=N$ and $b=b'=b''=N$.
Note that the first term between the curly brackets corresponds to
the planar-box two-meson-exchange diagram and the second term to the
crossed-box two-meson-exchange diagram. In these diagrams only the
contribution of the positive-energy nucleon states are included,
in accordance with the pair-suppression hypothesis that we used in
our earlier work on two-meson exchange~\cite{Rij91,Rij92a,Rij92b}.

The fourth-order potential $V^{(4)}$ consists of two parts. The first
part is given by the fourth-order planar and crossed BW diagrams, shown
in Figs.~\ref{pap1fig1}(a) and \ref{pap1fig1}(b)--(d), respectively,
and their ``mirror'' counterparts.
The second part is due to the fact that we do not distinguish the two
different time orderings (i.e., the explicit energy dependence) in the
one-boson-exchange diagrams. Hence, solving the scattering equation
does not generate the time-ordered TMO diagrams of Figs.~\ref{pap1fig2}(a)
and \ref{pap1fig2}(b), nor their ``mirror'' counterparts. We therefore
include these TMO diagrams explicitly~\cite{Fri94}, subtracting the
once-iterated one-meson contribution
\begin{equation}
  K^{(4)}_{{\rm Born}} = K^{(2)} g\ K^{(2)} =
          V^{(2)} g\ V^{(2)},                           \label{poteq1}
\end{equation}
and so the fourth-order potential reads
\begin{equation}
  V({\bf p}',{\bf p})=K^{(4)}-K^{(2)}g\ K^{(2)}.        \label{poteq2}
\end{equation}

The procedure to derive the kernels for the planar and crossed graphs
corresponding to Figs.~\ref{pap1fig1} and \ref{pap1fig2} is amply
described in Refs.~\cite{Rij92a,Rij92b} and will not be repeated here.
{}From the worked-out vertices (see below) and closely following the
procedure as discussed in the referred papers, the contribution of the
different graphs can be written down immediately.
Finally, when the mesons are distinguishable, we also have to
include the contributions from the diagrams where we interchange
the two meson lines (dashed and dotted lines in Figs.~\ref{pap1fig1}
and \ref{pap1fig2}). They merely give rise to a factor of two when
we evaluate the potentials.

For point couplings the nucleon-nucleon-meson Hamiltonians
are~\cite{Bjo65}
\begin{mathletters}
\begin{eqnarray}
  &&{\cal H}_{PV}=\frac{f_{PV}}{m_{\pi}}
     \bar{\psi} \gamma_{5}\gamma_{\mu}\bbox{\tau}\psi
     \!\cdot\!\partial^{\mu}\bbox{\phi}_{PV},          \label{Lagpv}\\
  &&{\cal H}_{S}=g_{S}\bar{\psi}\bbox{\tau}\psi
     \!\cdot\!\bbox{\phi}_{S},                          \label{Lags}\\
  &&{\cal H}_{V}= g_{V}\bar{\psi}\gamma_{\mu}\bbox{\tau}
     \psi\!\cdot\!\bbox{\phi}^{\mu}_{V} - \frac{f_{V}}{2M}
     \bar{\psi}\sigma_{\mu\nu}\bbox{\tau}\psi\!\cdot\!\partial^{\nu}
     \bbox{\phi}^{\mu}_{V},                             \label{Lagv}
\end{eqnarray}
\end{mathletters}
where $\bbox{\phi}$ denotes the pseudoscalar-, scalar-, and
vector-meson field, respectively. For the isoscalar ($I=0$)
mesons, the isospin Pauli matrices, $\bbox{\tau}$, are absent.
The nucleon-nucleon-Pomeron vertex has the same Lorentz structure
as the isoscalar scalar vertex.

As is well known, the pseudoscalar one-meson exchange can also be
represented by the pseudoscalar interaction Hamiltonian
\begin{equation}
   {\cal H}_{PS}=g_{PS}\bar{\psi}i\gamma_{5}\bbox{\tau}\psi
         \!\cdot\!\bbox{\phi}_{PS}.                    \label{Lagps}
\end{equation}
Both ${\cal H}_{PV}$ and ${\cal H}_{PS}$ give rise to the same
on-energy-shell one-meson-exchange potential provided that the
coupling constants satisfy the relation $f_{PV}/m_{\pi}=g_{PS}/2M$.
{}From this we tacitly assume that the pseudovector coupling constant
$f_{PV}$ is ${\cal O}(1/M)$. This simplifies the expressions for the
pion-pseudoscalar potentials considerably.

The generalization of the interaction kernels to the case with a
Gaussian (or any other) form factor has been treated and explained
in \cite{Rij91}.
The same procedure can be applied to the various meson exchanges to
be considered here. The form factors $F_{\alpha}({\bf k}^{2})$ and
$F_{\pi}({\bf k}^{2})$, which describe the meson-exchange and
$\pi$-exchange amplitudes, respectively, are simply a product of the
Gaussian vertex form factors $F_{N\!N\!\alpha}({\bf k}^{2})=
\exp(-{\bf k}^{2}/2\Lambda^{2}_{N\!N\!\alpha})$ and
$F_{N\!N\!\pi}({\bf k}^{2})=\exp(-{\bf k}^{2}/2\Lambda^{2}_{N\!N\!\pi})$.

\section{DEFINITION OF NUCLEON-NUCLEON \protect\\ TME POTENTIAL}
\label{sec:chap3}
The transition from Dirac spinors to Pauli spinors is reviewed in
Appendix C of Ref.~\cite{Rij91}. There we derived the
Lippmann-Schwinger equation,
\begin{eqnarray}
  \chi({\bf p}')&=&\chi^{(0)}({\bf p}')+\tilde{g}({\bf p}')\int\!d^{3}p\
  {\cal V}({\bf p}',{\bf p})\ \chi({\bf p}),               \label{LSeq2}
\end{eqnarray}
for the Pauli-spinor wave functions $\chi({\bf p})$. The wave function
$\chi({\bf p})$ and the potential ${\cal V}({\bf p}',{\bf p})$ in the
Pauli-spinor space are defined by
\begin{equation}
  \phi({\bf p})=\sum_{\sigma_{a},\sigma_{b}}
    \chi_{\sigma_{a}\sigma_{b}}({\bf p})\
    u_{a}({\bf p},\sigma_{a})u_{b}(-{\bf p},\sigma_{b})
\end{equation}
and
\begin{eqnarray}
  \chi^{(a)\dagger}_{\sigma'_{a}}\chi^{(b)\dagger}_{\sigma'_{b}}\;
  {\cal V}\; \chi^{(a)}_{\sigma_{a}}\chi^{(b)}_{\sigma_{b}}&=&
  \bar{u}_{a}({\bf p}',\sigma'_{a})\bar{u}_{b}(-{\bf p}',\sigma'_{b})
            V({\bf p}',{\bf p})                              \nonumber\\
  &\times& u_{a}({\bf p},\sigma_{a})u_{b}(-{\bf p},\sigma_{b}).
                                                         \label{LSitems}
\end{eqnarray}

Like in the derivation of the OBE potentials~\cite{Nag78,Mae89},
we make the approximation
\[
      E({\bf p})=({\bf p}^{2}+M^{2})^{1/2}\approx\ M+{\bf p}^{2}/2M
\]
everywhere in the interaction kernels of Sec.~\ref{sec:chap4}, which of
course is fully justified for low energies only. We have a similar
expansion for the on-shell energy,
\[
      W=2({\bf p}_{i}^{2}+M^{2})^{1/2}\approx\ 2M+{\bf p}_{i}^{2}/M.
\]
In contrast to this kind of approximation, the full ${\bf k}^{2}$
dependence of the form factors is kept throughout the derivation of the
two-meson-exchange potential.
Note that the Gaussian form factors suppress the high-momentum transfers
strongly. This means that the contribution to the potentials from
intermediate states which are far off-energy-shell cannot be very large.
It also means that the $1/M$ expansions can be expected to be valid
up to larger energies as well.

The reduction of the TME potential from Dirac-spinor space to Pauli-spinor
space is completely similar to the procedures discussed in
Refs.~\cite{Rij91,Rij92a,Rij92b}. The vertex operators in Pauli-spinor
space up to order $1/M^{2}$ are given by
\widetext
\begin{eqnarray}
  \bar{u}({\bf p}')\Gamma^{(a)}_{PV}u({\bf p})
  &=& -i\left(\frac{f_{PV}}{m_{\pi}}\right)
      \left[\bbox{\sigma}_{1}\!\cdot\!{\bf k}
        \left(1-\frac{{\bf p}^{\prime2}+{\bf p}^{2}}{8M^{2}}\right)
      \pm \frac{\omega}{2M}\bbox{\sigma}_{1}\!\cdot\!({\bf p}'+{\bf p})
      + \frac{{\bf p}^{\prime2}-{\bf p}^{2}}{8M^{2}}
        \bbox{\sigma}_{1}\!\cdot\!({\bf p}'+{\bf p})\right],
\nonumber\\ && \label{NNPV}\\
  \bar{u}({\bf p}')\Gamma^{(a)}_{PS}u({\bf p})
  &=& -i\frac{g_{PS}}{2M} \left[\bbox{\sigma}_{1}\!\cdot\!{\bf k}
        \left(1-\frac{{\bf p}^{\prime2}+{\bf p}^{2}}{8M^{2}}\right)
      - \frac{{\bf p}^{\prime2}-{\bf p}^{2}}{8M^{2}}
     \bbox{\sigma}_{1}\!\cdot\!({\bf p}'+{\bf p})\right], \label{NNPS}\\
 \bar{u}({\bf p}')\Gamma^{(a)}_{S}u({\bf p})
   &=& g_{S} \left[1-\frac{{\bf p}'\!\cdot\!{\bf p}
         +i\bbox{\sigma}_{1}\!\cdot\!{\bf p}'\times{\bf p}}{4M^{2}}
              \right],                                     \label{NNS}\\
  \bar{u}({\bf p}')\Gamma^{(a)}_{V}u({\bf p})
  &=& g_{V}\left[1+\frac{{\bf p}'\!\cdot\!{\bf p}
               +i\bbox{\sigma}_{1}\!\cdot\!{\bf p}'\times{\bf p}}{4M^{2}}
      - \kappa_{V}\frac{({\bf p}'-{\bf p})^{2}-2i\bbox{\sigma}_{1}
               \!\cdot\!{\bf p}'\times{\bf p}}{4M^{2}} \right]
      \phi_{V}^{0}                                         \nonumber\\
  &-& \frac{g_{V}}{2M} \biggl[ ({\bf p}'+{\bf p})
         +i(1+\kappa_{V})\bbox{\sigma}_{1}\times{\bf k}\biggr]
         \cdot\bbox{\phi}_{V},                             \label{NNV}
\end{eqnarray}
\narrowtext \noindent
where ${\bf k}={\bf p}'-{\bf p}$ and $\kappa_{V}=f_{V}/g_{V}$.
The expressions for $\bar{u}(-{\bf p}')\Gamma^{(b)}u(-{\bf p})$ are
trivially obtained by substituting
$({\bf p}',{\bf p},{\bf k},\bbox{\sigma}_{1})\rightarrow
(-{\bf p}',-{\bf p},-{\bf k},\bbox{\sigma}_{2})$.
For the isovector ($I=1$) mesons, the appropriate isospin matrices have
to be attached. In the $\Gamma$-matrix element of Eq.~(\ref{NNPV}),
the upper sign applies to the creation and the lower sign to the
absorption of the pseudoscalar meson at the vertex.

Although we will assume the pseudovector coupling for the pion,
we here have also listed the vertex operator in case of pseudoscalar
coupling for reasons of completeness, which might be helpful for
later work by other groups. The changes can be easily accounted for,
as we will indicate in the following sections.

Useful for the evaluation of the second-order diagrams are the relations
\begin{eqnarray}
  \tau_{j}\tau_{i}=\delta_{ij}+i\epsilon_{jik}\tau_{k}, &\hspace{5ex}&
  \sigma_{j}\sigma_{i}=\delta_{ij}+i\epsilon_{jik}\sigma_{k}.
\end{eqnarray}
Products of this type will occur for each nucleon line.
The isospin factors for the planar (//) and crossed ($X$) diagrams can
readily be evaluated. In the case of two isospin-1 mesons, we have
\begin{eqnarray}
  C^{(//)}_{N\!N}(I) &=&
          3-2\bbox{\tau}_{1}\!\cdot\!\bbox{\tau}_{2},       \nonumber\\
  C^{(X)}_{N\!N}(I) &=&
          3+2\bbox{\tau}_{1}\!\cdot\!\bbox{\tau}_{2},      \label{CNN1}
\end{eqnarray}
where $I$ denotes the total isospin of the $N\!N$ state. For the
exchange of an isospin-1 with an isospin-0 exchange, one has
\begin{equation}
  C^{(//)}_{N\!N}(I)=C^{(X)}_{N\!N}(I)\equiv C^{(1,0)}_{N\!N}(I)=
      \bbox{\tau}_{1}\!\cdot\!\bbox{\tau}_{2}.             \label{CNN0}
\end{equation}

In order to obtain the contributions to the potentials in the
adiabatic approximation, we expand the energy denominators in
the expressions for the planar and the crossed-box diagrams
and keep only the leading term. So, for example,
\begin{eqnarray}
   \frac{1}{E({\bf p})+{E}({\bf p}'')-W\!+\omega}
      &\approx&  \frac{1}{\omega} \left[ 1 +
         \frac{{\bf k}_{1}\!\cdot\!{\bf k}_{2}\!-\!
              {\bf q}\!\cdot\!({\bf k}_{1}\!-\!{\bf k}_{2})}{2M\omega}
         \right], \nonumber\\
   & & \label{approx2}
\end{eqnarray}
where we have made the on-energy-shell approximation
${\bf p}^{2}-{\bf p}^{2}_{i}=0$ and ${\bf p}^{\prime2}-{\bf p}^{2}_{i}=0$,
with ${\bf p}_{i}$ the initial-state momentum, and where we introduced
${\bf q}={\textstyle\frac{1}{2}}({\bf p}'+{\bf p})$. The second term
between square brackets gives the nonadiabatic contribution to the
potentials, presented in Sec.~\ref{sec:chap5a}. The term proportional
to ${\bf q}\!\cdot\!({\bf k}_{1}-{\bf k}_{2})$ usually drops out for
symmetry reasons. The only exceptions occur for the planar-box
nonadiabatic pion-pseudoscalar potentials (see Sec.~\ref{sec:chap5a}).

In the following, we will concentrate on the derivation of the
pion-meson potentials, which are expected to be the most important
ones. The derivation of the more general case of two arbitrary mesons
(not necessarily involving at least one pion) will then be straight
forward.

\section{PION-MESON-EXCHANGE POTENTIALS}
\label{sec:chap4}
\subsection{Pion-pseudoscalar exchange}
\label{sec:chap4a}
Using the assumption that the pseudovector coupling constants are
${\cal O}(1/M)$ [see the discussion below Eq.~(\ref{Lagpv})], we can
neglect all $1/M^{2}$ contributions and the pion-pseudoscalar-exchange
potential is the simplest potential we will encounter.
As a representative of the pseudoscalar mesons we select the $\eta$
and derive the $\pi\eta$-exchange potential. The $\pi\pi$-exchange
potential has already been discussed in Ref.~\cite{Rij91}. The purely
off-shell term proportional to $({\bf p}^{\prime2}-{\bf p}^{2})$ in
the pseudovector vertex (\ref{NNPV}) gives rise to a $1/M$ term in the
TMO diagrams, rather than a $1/M^{2}$ term. We defer this contribution
to the potential to Sec.~\ref{sec:chap5c}. Similarly, the contribution
due to the $\omega/M$ term will be discussed in Sec.~\ref{sec:chap5b}.

We assign (${\bf k}_{1}, \omega_{1}$) to the $\pi$ meson and
(${\bf k}_{2}, \omega_{2}$) to the $\eta$ meson.
The planar BW graph of Fig.~\ref{pap1fig1}(a), the TMO graphs of
Fig.~\ref{pap1fig2}, and their ``mirror'' counterparts give
\widetext
\begin{eqnarray}
  V^{(0)}_{\pi\eta}(//) &=& C_{N\!N}^{(1,0)}(I)
    \left(\frac{f_{N\!N\!\eta}}{m_{\pi}}\right)^{2}
    \left(\frac{f_{N\!N\!\pi}}{m_{\pi}}\right)^{2}
    \int\!\!\int\!\frac{d^{3}k_{1}d^{3}k_{2}}{(2\pi)^{6}}\,
    e^{i({\bf k}_{1}+{\bf k}_{2})\cdot{\bf r}}\,
    F_{\pi}({\bf k}_{1}^{2})F_{\eta}({\bf k}_{2}^{2})        \nonumber\\
  &\times& \biggl[{\bf k}_{1}\!\cdot\!{\bf k}_{2}-
    i\bbox{\sigma}_{1}\!\cdot\!({\bf k}_{1}\times{\bf k}_{2})\biggr]
           \biggl[{\bf k}_{1}\!\cdot\!{\bf k}_{2}-
    i\bbox{\sigma}_{2}\!\cdot\!({\bf k}_{1}\times{\bf k}_{2})\biggr]
           D_{//}(\omega_{1},\omega_{2}),                  \label{pseu1}
\end{eqnarray}
and the crossed BW graphs of Figs.~\ref{pap1fig1}(b)--(d) and their
``mirror'' graphs give
\begin{eqnarray}
  V^{(0)}_{\pi\eta}(X) &=& C_{N\!N}^{(1,0)}(I)
    \left(\frac{f_{N\!N\!\eta}}{m_{\pi}}\right)^{2}
    \left(\frac{f_{N\!N\!\pi}}{m_{\pi}}\right)^{2}
    \int\!\!\int\!\frac{d^{3}k_{1}d^{3}k_{2}}{(2\pi)^{6}}\,
    e^{i({\bf k}_{1}+{\bf k}_{2})\cdot{\bf r}}\
    F_{\pi}({\bf k}_{1}^{2})F_{\eta}({\bf k}_{2}^{2})      \nonumber\\
  &\times& \biggl[{\bf k}_{1}\!\cdot\!{\bf k}_{2}-
    i\bbox{\sigma}_{1}\!\cdot\!({\bf k}_{1}\times{\bf k}_{2})\biggr]
           \biggl[{\bf k}_{1}\!\cdot\!{\bf k}_{2}+
    i\bbox{\sigma}_{2}\!\cdot\!({\bf k}_{1}\times{\bf k}_{2})\biggr]
           D_{X}(\omega_{1},\omega_{2}).                  \label{pseu2}
\end{eqnarray}
\narrowtext \noindent
Here $F_{\pi}({\bf k}_{1}^{2})=F^{2}_{N\!N\!\pi}({\bf k}_{1}^{2})$
and $F_{\eta}({\bf k}_{2}^{2})=F^{2}_{N\!N\!\eta}({\bf k}_{2}^{2})$,
and the energy denominators $D_{i}(\omega_{1},\omega_{2})$ are
given in Table~\ref{table1}.
Choosing the pseudoscalar coupling (\ref{NNPS}) instead of the
pseudovector coupling (\ref{NNPV}) will result in the same
leading-order potential.

The full separation of the ${\bf k}_{1}$ and ${\bf k}_{2}$ dependence
of the Fourier integrals can be achieved as shown in Ref.~\cite{Rij91}.
Using the $\lambda$-integral representation, Eq.~(B.4) of \cite{Rij91},
one can derive that
\begin{equation}
  \frac{1}{\omega^{3}} =
     \frac{2}{\pi}\int_{0}^{\infty}\!\frac{d\lambda}{\lambda^{2}}
    \left[\frac{1}{\omega^{2}}-\frac{1}{\omega^{2}+\lambda^{2}}\right],
                                                          \label{basic1}
\end{equation}
and
\begin{eqnarray}
   \frac{1}{\omega_{1}^{2}\omega_{2}^{2}}\frac{1}{\omega_{1}+\omega_{2}}
     = \frac{2}{\pi}\int_{0}^{\infty}\!\frac{d\lambda}{\lambda^{2}} &&
 \left[\frac{1}{\omega_{1}^{2}}-\frac{1}{\omega_{1}^{2}+\lambda^{2}}
       \right]                                               \nonumber\\
   \times&& \left[\frac{1}{\omega_{2}^{2}}-
        \frac{1}{\omega_{2}^{2}+\lambda^{2}}\right].      \label{basic2}
\end{eqnarray}
In performing the Fourier transform, the $1/\omega^{2}$ term gives
rise to the basic function $I_{2}(m,r)=(m/4\pi)\phi^{0}_{C}(m,r)$,
where~\cite{Nag78,Mae89}
\begin{eqnarray*}
   \phi^{0}_{C}(m,r) = e^{m^{2}/\Lambda^{2}} &&\left[e^{-mr} {\rm erfc}
           \left(\frac{m}{\Lambda}-\frac{\Lambda r}{2}\right) \right. \\
   & &\ \left. -e^{mr}{\rm erfc}\left(\frac{m}{\Lambda}+
               \frac{\Lambda r}{2}\right)\right]\frac{1}{2mr}.
\end{eqnarray*}
Similarly, the $1/(\omega^{2}+\lambda^{2})$ term gives rise to the
function
\begin{equation}
   F_{\alpha}(\lambda,r)=e^{-\lambda^{2}/\Lambda_{\alpha}^{2}}
          I_{2}(\sqrt{m_{\alpha}^{2}+\lambda^{2}},r).    \label{basicF}
\end{equation}
Therefore, the typical coordinate-space function for the planar and
crossed diagrams is of the form
\begin{eqnarray}
  B_{\alpha\beta}(r_{1},r_{2}) = \frac{2}{\pi}
       \int_{0}^{\infty}\!\frac{d\lambda}{\lambda^{2}}  &&
       \biggl[I_{2}(m_{\alpha},r_{1})I_{2}(m_{\beta},r_{2})  \nonumber\\
  &&     -F_{\alpha}(\lambda,r_{1})F_{\beta}(\lambda,r_{2})\biggr].
                                                             \label{B22}
\end{eqnarray}
Accordingly, we derive from Eqs.~(\ref{pseu1}) and (\ref{pseu2}) the form
\begin{eqnarray}
 V^{(0)}_{\pi\eta} &=& \sum_{i=//,X} C_{N\!N}^{(1,0)}(I)
      \left(\frac{f_{N\!N\!\eta}}{m_{\pi}}\right)^{2}
      \left(\frac{f_{N\!N\!\pi}}{m_{\pi}}\right)^{2}
      \lim_{{\bf r}_{1},{\bf r}_{2}\rightarrow{\bf r}}       \nonumber\\
  &\times& O^{(i)}_{PS}(-i\bbox{\nabla}_{1},-i\bbox{\nabla}_{2})\
     B_{N\!N\!\pi\eta}^{(i)}(r_{1},r_{2}),                 \label{potps}
\end{eqnarray}
where
\begin{equation}
   B^{(//)}_{N\!N\!\pi\eta}(r_{1},r_{2}) =
  -B^{(X)}_{N\!N\!\pi\eta}(r_{1},r_{2}) = {\textstyle\frac{1}{2}}
     B_{\pi\eta}(r_{1},r_{2}),                           \label{BNNfunc}
\end{equation}
and the operators $O^{(i)}_{PS}({\bf k}_{1},{\bf k}_{2})$ are given
in Table~\ref{table2}. We should point out that these operators contain
the contributions from {\it all} graphs, including the contributions
where we interchanged the $\pi$ and $\eta$ lines. This gives an
overcounting when the particles are identical, and so in case of
$\pi\pi$ exchange these operators should be divided by 2.

\subsection{Pion-scalar exchange}
\label{sec:chap4b}
To be definite we select from the scalar mesons the $a_{0}(980)$ and
derive the $\pi a_{0}$-exchange potential. The $\pi\varepsilon(760)$-
and the $\pi f_{0}(975)$-exchange potentials can simply be obtained by
an appropriate change of the isospin structure. In the following,
(${\bf k}_{1}, \omega_{1}$) refers to the $\pi$ meson and (${\bf k}_{2},
\omega_{2}$) refers to the $a_{0}$ meson.

We will evaluate the potentials up to order $1/M^{2}$.
Using the vertices of Eqs.~(\ref{NNPV}) and (\ref{NNS}), and including
the normalization factors of the nucleon Dirac spinors in the
intermediate two-nucleon state, which also contribute a factor of order
$1/M^{2}$, we readily obtain the contributions to the $N\!N$ potentials.
After some rearrangement, the planar graphs give
\widetext
\begin{eqnarray}
  V^{(0)}_{\pi a_{0}}(//) &=& C_{N\!N}^{(//)}(I)
    g_{N\!N a_{0}}^{2} \left(\frac{f_{N\!N\!\pi}}{m_{\pi}}\right)^{2}
    \int\!\!\int\!\frac{d^{3}k_{1}d^{3}k_{2}}{(2\pi)^{6}}\,
    e^{i({\bf k}_{1}+{\bf k}_{2})\cdot{\bf r}}\
    F_{\pi}({\bf k}_{1}^{2})F_{a_{0}}({\bf k}_{2}^{2})    \nonumber\\
  &\times&  \Biggl\{ \left[
     1-\frac{{\bf p}'\!\cdot\!{\bf p}''
        +i\bbox{\sigma}_{1}\!\cdot\!{\bf p}'\times{\bf p}''}{4M^{2}}
     \right] \left[
     1-\frac{{\bf p}'\!\cdot\!{\bf p}''
        +i\bbox{\sigma}_{2}\!\cdot\!{\bf p}'\times{\bf p}''}{4M^{2}}
     \right] (\bbox{\sigma}_{1}\!\cdot\!{\bf k}_{1})
             (\bbox{\sigma}_{2}\!\cdot\!{\bf k}_{1})
             N_{//}({\bf p}'')                             \nonumber\\
  && \ + \frac{{\bf p}^{\prime\prime2}-{\bf p}^{2}}{8M^{2}}
     \biggl[\bbox{\sigma}_{1}\!\cdot\!{\bf k}_{1}
            \bbox{\sigma}_{2}\!\cdot\!({\bf p}''+{\bf p}) +
            \bbox{\sigma}_{1}\!\cdot\!({\bf p}''+{\bf p})
            \bbox{\sigma}_{2}\!\cdot\!{\bf k}_{1}\biggr] \Biggr\}
     D_{//}(\omega_{1},\omega_{2}),                     \label{scal1}
\end{eqnarray}
and the crossed graphs give
\begin{eqnarray}
  V^{(0)}_{\pi a_{0}}(X) &=& C_{N\!N}^{(X)}(I)
    g_{N\!N a_{0}}^{2} \left(\frac{f_{N\!N\!\pi}}{m_{\pi}}\right)^{2}
    \int\!\!\int\!\frac{d^{3}k_{1}d^{3}k_{2}}{(2\pi)^{6}}\,
    e^{i({\bf k}_{1}+{\bf k}_{2})\cdot{\bf r}}\
    F_{\pi}({\bf k}_{1}^{2})F_{a_{0}}({\bf k}_{2}^{2})    \nonumber\\
  &\times&  \Biggl\{ \left[
     1-\frac{{\bf p}'\!\cdot\!{\bf p}''
        +i\bbox{\sigma}_{1}\!\cdot\!{\bf p}'\times{\bf p}''}{4M^{2}}
     \right]  (\bbox{\sigma}_{1}\!\cdot\!{\bf k}_{1})
              (\bbox{\sigma}_{2}\!\cdot\!{\bf k}_{1})  \left[
     1-\frac{{\bf p}'''\!\cdot\!{\bf p}
        +i\bbox{\sigma}_{2}\!\cdot\!{\bf p}'''\times{\bf p}}{4M^{2}}
     \right] N_{X}({\bf p}'',{\bf p}''')                 \nonumber\\
  && \ + \frac{{\bf p}^{\prime2}-{\bf p}^{\prime\prime\prime2}}{8M^{2}}
           \bbox{\sigma}_{1}\!\cdot\!{\bf k}_{1}
           \bbox{\sigma}_{2}\!\cdot\!({\bf p}'+{\bf p}''')
       + \frac{{\bf p}^{\prime\prime2}-{\bf p}^{2}}{8M^{2}}
           \bbox{\sigma}_{1}\!\cdot\!({\bf p}''+{\bf p})
           \bbox{\sigma}_{2}\!\cdot\!{\bf k}_{1}        \Biggr\}
   D_{X}(\omega_{1},\omega_{2}).                       \label{scal2}
\end{eqnarray}
\narrowtext \noindent
Here we introduced momentum vectors for the intermediate nucleon lines,
given by
\begin{eqnarray}
   {\bf p}'' &=& {\bf p}+{\bf k}_{1}={\bf p}'-{\bf k}_{2},   \nonumber\\
   {\bf p}'''&=& {\bf p}+{\bf k}_{2}={\bf p}'-{\bf k}_{1},
\end{eqnarray}
and the functions
\begin{eqnarray}
   && N_{//}({\bf p}'')=1
       +\frac{{\bf p}^{\prime\prime2}-{\bf p}^{2}}{4M^{2}}, \nonumber\\
   && N_{X}({\bf p}'',{\bf p}''')=1
     +\frac{{\bf p}^{\prime\prime2}-{\bf p}^{2}}{8M^{2}}
     +\frac{{\bf p}^{\prime\prime\prime2}-{\bf p}^{\prime2}}{8M^{2}}.
\end{eqnarray}
Again, the energy denominators $D_{i}(\omega_{1},\omega_{2})$ are
given in Table~\ref{table1}, and for all graphs
$F_{\pi}({\bf k}_{1}^{2})=F^{2}_{N\!N\!\pi}({\bf k}_{1}^{2})$ and
$F_{a_{0}}({\bf k}_{2}^{2})=F^{2}_{N\!N a_{0}}({\bf k}_{2}^{2})$.
Including the ``time-reversed'' diagrams, the resulting coordinate-space
potential can be written in the form
\begin{eqnarray}
 V^{(0)}_{\pi a_{0}} &=& \sum_{i=//,X} C_{N\!N}^{(i)}(I)
   g_{N\!N a_{0}}^{2} \left(\frac{f_{N\!N\!\pi}}{m_{\pi}}\right)^{2}
   \lim_{{\bf r}_{1},{\bf r}_{2}\rightarrow{\bf r}}          \nonumber\\
   &\times& O^{(i)}_{S}(-i\bbox{\nabla}_{1},-i\bbox{\nabla}_{2})\
     B_{N\!N\pi a_{0}}^{(i)}(r_{1},r_{2}),              \label{pots}
\end{eqnarray}
where $B_{N\!N\pi a_{0}}^{(i)}(r_{1},r_{2})$ is defined by
Eq.~(\ref{BNNfunc}).
The operators $O^{(i)}_{S}({\bf k}_{1},{\bf k}_{2})$ can be found
in Table~\ref{table2}. They contain the contributions from {\it all}
graphs. The choice of pseudoscalar coupling (\ref{Lagps}) for the
pion instead of pseudovector coupling (\ref{Lagpv}) results in a
change of sign in one of the terms, as indicated in Table~\ref{table2}.

The purely off-shell contribution proportional to
$({\bf p}^{\prime\prime2}-{\bf p}^{2})$ gives rise to a $1/M$ term in
the TMO diagrams, which will be discussed in Sec.~\ref{sec:chap5c}.
Similarly, the contribution due to the $\omega/M$ term in the
pion vertex (\ref{NNPV}) will be discussed separately in
Sec.~\ref{sec:chap5b}.

Note that, up to this order, both planar and crossed diagrams give
rise to the same momentum operators $O_{S}({\bf k}_{1},{\bf k}_{2})$.
As a consequence, the planar and crossed contributions for the
$\pi\varepsilon$ and $\pi f_{0}$ potentials exactly cancel, and hence
\begin{equation}
    V^{(0)}_{\pi\varepsilon} = V^{(0)}_{\pi f_{0}} = 0.
\end{equation}

\subsection{Pion-Pomeron exchange}
\label{sec:chap4c}
The pion-Pomeron potential has the same structure as the pion-scalar
potential, except for an over-all minus sign. Hence,
\begin{eqnarray}
            V^{(0)}_{\pi P} = 0.
\end{eqnarray}

Although the leading-order part of the potential vanishes, the $1/M$
corrections to the potential do not, and so in order to be able to
evaluate the pion-Pomeron potentials in coordinate space, it is
appropriate here first to point out the peculiarities of Pomeron
exchange with respect to meson exchange.
We use as a working hypothesis that there is a contribution to the
potential which corresponds to the Pomeron phenomenon at high energies.
In QCD, the physical nature of Pomeron exchange is understood as
color-singlet two- or multigluon exchange~\cite{Low75,Nus75,Sim90}.
The assumption that the Pomeron for our purposes can be described
effectively by a simple Regge pole is part of our phenomenological
and practical approach. In the Nijmegen OBE models Pomeron exchange
provides a significant contribution to the short-range repulsion. We
refer the interested reader to Refs.~\cite{Nag78,Mae89,Rij75}, and
references cited therein. It remains to be seen whether Pomeron
exchange, using some plausible form for it, is still a useful concept
when we include many more contributions (i.e., two-meson contributions)
than were taken into account in the Nijmegen OBE models.
But the inclusion of the potentials involving the Pomeron is important
in order to have the full consequences of the Pomeron worked out.
Hence, at this stage these potentials are important for an extension
of the Nijmegen OBE model.

We treat the Pomeron as a Regge pole, using the Durand-van Hove
model~\cite{Dur67}. Accordingly, in dealing with Pomeron exchange,
we replace the OBE factor for meson exchange by an infinite sum over
all even $J$, i.e.
\begin{equation}
    \frac{g^{2}}{m^{2}-t} \Rightarrow A_{P}(t) = \sum_{J\ {\rm even}}
    (2J+1) \frac{g^{2}(J)}{\omega_{J}^{2}(t)},            \label{pom1}
\end{equation}
where
\begin{equation}
    \omega_{J}^{2}(t)= m_{J}^{2}-t \approx {\bf k}^{2}+m_{J}^{2}.
    \label{pom1a}
\end{equation}
Following the same procedure as in the OBE model~\cite{Rij75}
(i.e., the Khuri-Jones procedure for a Regge pole), and using the
Sommerfeld-Watson transform, one gets for the Durand-van Hove sum
in Eq.~(\ref{pom1}) the expression
\begin{eqnarray}
   A_{P}(t) &=& \sum_{J\ {\rm even}} (2J+1) A_{J}(t),     \nonumber \\
   A_{J}(t) &=&-\beta(t)\frac{e^{-[J-\alpha(t)]\xi}}{\alpha(t)-J}.
                                                          \label{pom2}
\end{eqnarray}
Here $\alpha(t)$ is the solution of the equation $m_{J}^{2}-t=0$,
$\beta(t)$ is the residue at the pole, and $\xi$ is a parameter in the
Khuri-Jones representation (see \cite{Rij75}). For low energies, our
region of interest here, $\xi$ is such that one can restrict oneself to
the lowest $J$ value in Eq.~(\ref{pom2}), which is $J=0$.
However, we should point out that in the case of the Pomeron the
$J=0$ contribution from the Khuri-Jones procedure has nothing to do
with a physical particle.
With the Chew ghost-killing factor, $\beta(t)=-\alpha(t)\exp(a_{P}t)$,
one obtains for Pomeron exchange at low energies from
Eq.~(\ref{pom2}) the representation~\cite{Rij75}
\begin{equation}
    A_{P}(t) = \frac{g^{2}_{P}}{M^{2}}
               \exp \left(\frac{t}{4m_{P}^{2}}\right), \label{pom3}
\end{equation}
where the Pomeron-nucleon-nucleon coupling and the effective
Pomeron mass can be expressed as~\cite{Rij75}
\begin{eqnarray}
    g^{2}_{P} &=& \gamma_{P}(0)
             \left(\bar{s}_{R}/M^{2}\right)^{\alpha_{P}(0)}, \nonumber\\
    m_{P}^{2} &=& \frac{1}{4}\left[a_{P}+\alpha'_{P}\ln
           \frac{\bar{s}_{R}}{M^{2}}\right]^{-1}.           \label{pom4}
\end{eqnarray}

With this preparation, we can now explain the treatment of the
pion-Pomeron contribution to the potentials. As we will see in
the next section, for the Pomeron contributions we encounter in
the energy denominators the factors $1/\omega_{J}^{2}$ and
$1/\omega_{J}^{4}$. The handling of the first term is discussed above
and that of the second is easily inferred by observing that
\begin{equation}
  A_{4}(t)=\sum_{J\ {\rm even}}(2J+1)\frac{g^{2}(J)}{\omega_{J}^{4}(t)}
      =\frac{d}{dt}A_{P}(t).                               \label{pom5}
\end{equation}
{}From Eq.~(\ref{pom3}) we therefore find that
\begin{equation}
     A_{4}(t) = \frac{1}{4m_{P}^{2}}A_{P}(t).              \label{pom6}
\end{equation}
The result of this analysis is that the basic Pomeron functions are
\begin{eqnarray}
     I_{2}(m,r) &\rightarrow& I_{G}(m_{P},r),            \nonumber \\
     I_{4}(m,r) &\rightarrow& \frac{1}{4m_{P}^{2}} I_{G}(m_{P},r),
                                                           \label{pom7}
\end{eqnarray}
where
\begin{equation}
    I_{G}(m_{P},r) = \frac{m_{P}}{4\pi} \frac{4}{\sqrt{\pi}}
    \left(\frac{m_{P}}{M}\right)^{2}e^{-m_{P}^{2}r^{2}}. \label{pom8}
\end{equation}
With the substitutions (\ref{pom7}), the treatment of the Pomeron
is now completely analogous to that of the mesons.

\subsection{Pion-vector exchange}
\label{sec:chap4d}
Due to the complexity of the vector-meson vertices (\ref{NNV}),
it is not very illuminating to give the various intermediate steps
in the evaluation of the pion-vector-exchange potential.
It suffices to note that it is convenient to separate the potential
in four parts proportional to 1, $(1+\kappa_{V})$, $(1+2\kappa_{V})$,
and $(1+\kappa_{V})^{2}$, respectively.
We refer to these as the electric-electric ($e,e$), the two
electric-magnetic ($e,m$) and ($e,2m$), and the magnetic-magnetic
($m,m$) terms.
Assigning again (${\bf k}_{1}, \omega_{1}$) to the $\pi$ meson
and (${\bf k}_{2}, \omega_{2}$) to the $\rho$ meson, we can then
write the resulting potential as
\begin{eqnarray}
 V^{(0)}_{\pi\!\rho} &=& \sum_{i=//,X} C_{N\!N}^{(i)}(I)
   g_{N\!N\!\rho}^{2} \left(\frac{f_{N\!N\!\pi}}{m_{\pi}}\right)^{2}
                                                             \nonumber\\
 &\times& \int\!\!\int\!\frac{d^{3}k_{1} d^{3}k_{2}}{(2\pi)^{6}}\,
   e^{i({\bf k}_{1}+{\bf k}_{2})\cdot{\bf r}}
   F_{\pi}({\bf k}_{1}^{2}) F_{\rho}({\bf k}_{2}^{2})        \nonumber\\
 &\times& \biggl[ \frac{1}{2M^{2}}\biggl\{ O^{(i)}_{e,m}+O^{(i)}_{e,2m}
     +O^{(i)}_{m,m}\biggr\}({\bf k}_{1},{\bf k}_{2})         \nonumber\\
 & & + O^{(i)}_{e,e}({\bf k}_{1},{\bf k}_{2}) \biggr]
   D_{i}(\omega_{1},\omega_{2}),                         \label{vec1}
\end{eqnarray}
where the operators $O^{(i)}({\bf k}_{1},{\bf k}_{2})$ can be
found in Table~\ref{table2}. They contain the contributions from
{\it all} graphs. Again, the choice of pseudoscalar coupling for
the pion instead of pseudovector coupling involves a simple
change of sign in one of the terms in $O^{(i)}_{e,e}$.

The coordinate-space potential reads
\begin{eqnarray}
 V^{(0)}_{\pi\!\rho} &=& \sum_{i=//,X} C_{N\!N}^{(i)}(I)
   g_{N\!N\!\rho}^{2} \left(\frac{f_{N\!N\!\pi}}{m_{\pi}}\right)^{2}
   \lim_{{\bf r}_{1},{\bf r}_{2}\rightarrow{\bf r}}          \nonumber\\
 &\times& \biggl[ \frac{1}{2M^{2}}\biggl\{ O^{(i)}_{e,m}
   +O^{(i)}_{e,2m}+O^{(i)}_{m,m} \biggr\}
   (-i\bbox{\nabla}_{1},-i\bbox{\nabla}_{2})                \nonumber\\
 && + O^{(i)}_{e,e}(-i\bbox{\nabla}_{1},-i\bbox{\nabla}_{2}) \biggr]
    B^{(i)}_{N\!N\!\pi\!\rho}(r_{1},r_{2}),              \label{potv}
\end{eqnarray}
The $\pi\omega$ and $\pi\phi$ potentials are again obtained by an
appropriate change of the isospin structure.
Note that in leading order (no $1/M^{2}$ contributions), the planar
and crossed-box contributions exactly cancel for both $\pi\omega$
and $\pi\phi$ potentials; see Eq.~(\ref{B8}).

\section{1/M CORRECTIONS}
\label{sec:chap5}
\subsection{Nonadiabatic corrections}
\label{sec:chap5a}
The nonadiabatic correction from the $1/M$ expansion of the energy
denominators is explained in Ref.~\cite{Rij91}. The expansion of
the energy denominators involves a momentum dependence which for
all cases can be rewritten in the form $[{\bf k}_{1}\!\cdot\!{\bf k}_{2}
\pm{\bf q}\!\cdot\!({\bf k}_{1}-{\bf k}_{2})]/2M$.
Separating-off the momentum dependence, the energy denominators of
Table~\ref{table1} give rise to the nonadiabatic energy denominators
$D^{(1)}_{i}$ as listed in Table~\ref{table3}.
Note that $D^{(1)}_{X}=-2D^{(1)}_{//}$, and so now the contributions
from the planar and crossed diagrams for $\pi\varepsilon(760)$,
$\pi f_{0}(975)$, and $\pi P$ do not cancel.
We should point out that this expansion is only approximately valid,
and in principle breaks down at the pion-production threshold
($T_{\rm lab}\approx 280$ MeV).

In the crossed-box diagrams the ${\bf q}$ dependence always cancels.
Similarly, for pion-vector and pion-scalar (pion-Pomeron) exchange,
the interchange of the two meson lines also removes the ${\bf q}$
dependence. Hence, for pion-scalar exchange we get
\begin{eqnarray}
   V^{(1a)}_{\pi\varepsilon}(i) &=& C_{N\!N}^{(1,0)}(I)
     g_{N\!N\!\varepsilon}^{2}
     \left(\frac{f_{N\!N\!\pi}}{m_{\pi}}\right)^{2}
     \left(\frac{1}{M}\right)                                \nonumber\\
   &\times& \int\!\!\int\frac{d^{3}k_{1}d^{3}k_{2}}{(2\pi)^{6}}\,
     e^{i({\bf k}_{1}+{\bf k}_{2})\cdot{\bf r}}\
     F_{\pi}({\bf k}_{1}^{2})F_{\varepsilon}({\bf k}_{2}^{2})\nonumber\\
   &\times& ({\bf k}_{1}\!\cdot\!{\bf k}_{2})
     (\bbox{\sigma}_{1}\!\cdot\!{\bf k}_{1})
     (\bbox{\sigma}_{2}\!\cdot\!{\bf k}_{1})
     D_{i}^{(1)}(\omega_{1},\omega_{2}),                  \label{nonadS}
\end{eqnarray}
where we took the $\pi\varepsilon$ potential as a particular example.
Obviously, for pion-Pomeron exchange we get the same expression
as for pion-scalar exchange, except for an over-all minus sign.
For pion-vector exchange we get
\begin{eqnarray}
   V^{(1a)}_{\pi\!\rho}(i) &=& -C_{N\!N}^{(i)}(I)
     g_{N\!N\!\rho}^{2} \left(\frac{f_{N\!N\!\pi}}{m_{\pi}}\right)^{2}
     \left(\frac{1}{M}\right)                                \nonumber\\
   &\times& \int\!\!\int\frac{d^{3}k_{1}d^{3}k_{2}}{(2\pi)^{6}}\,
     e^{i({\bf k}_{1}+{\bf k}_{2})\cdot{\bf r}}\
     F_{\pi}({\bf k}_{1}^{2}) F_{\rho}({\bf k}_{2}^{2})      \nonumber\\
   &\times& ({\bf k}_{1}\!\cdot\!{\bf k}_{2})
     (\bbox{\sigma}_{1}\!\cdot\!{\bf k}_{1})
     (\bbox{\sigma}_{2}\!\cdot\!{\bf k}_{1})
     D_{i}^{(1)}(\omega_{1},\omega_{2}),                  \label{nonadV}
\end{eqnarray}
where we took the $\pi\!\rho$ potential as a particular example, and
only kept the leading term in $O_{e,e}^{(i)}({\bf k}_{1},{\bf k}_{2})$.

The situation for pion-pseudoscalar exchange is more subtle.
As mentioned above, in the crossed-box diagrams the ${\bf q}$
dependence cancels, resulting in
\widetext
\begin{eqnarray}
   V^{(1a)}_{\pi\eta}(X) &=& C_{N\!N}^{(1,0)}(I)
     \left(\frac{f_{N\!N\!\eta}}{m_{\pi}}\right)^{2}
     \left(\frac{f_{N\!N\!\pi}}{m_{\pi}}\right)^{2}
     \left(\frac{1}{M}\right)
     \int\!\!\int\frac{d^{3}k_{1}d^{3}k_{2}}{(2\pi)^{6}}\,
     e^{i({\bf k}_{1}+{\bf k}_{2})\cdot{\bf r}}\
     F_{\pi}({\bf k}_{1}^{2}) F_{\eta}({\bf k}_{2}^{2})      \nonumber\\
   &\times& ({\bf k}_{1}\!\cdot\!{\bf k}_{2})
     \biggl[ ({\bf k}_{1}\!\cdot\!{\bf k}_{2})^{2}
     +\bbox{\sigma}_{1}\!\cdot\!({\bf k}_{1}\times{\bf k}_{2})
      \bbox{\sigma}_{2}\!\cdot\!({\bf k}_{1}\times{\bf k}_{2}) \biggr]
     D_{X}^{(1)}(\omega_{1},\omega_{2}),                \label{nonadPVx}
\end{eqnarray}
where we took the $\pi\eta$ potential as a particular example.
In the planar-box diagram, on the other hand, the ${\bf q}$ dependence
survives and, after rearranging, we are left with
\begin{eqnarray}
   V^{(1a)}_{\pi\eta}(//) &=& C_{N\!N}^{(1,0)}(I)
     \left(\frac{f_{N\!N\!\eta}}{m_{\pi}}\right)^{2}
     \left(\frac{f_{N\!N\!\pi}}{m_{\pi}}\right)^{2}
     \left(\frac{1}{M}\right)
     \int\!\!\int\frac{d^{3}k_{1}d^{3}k_{2}}{(2\pi)^{6}}\,
     e^{i({\bf k}_{1}+{\bf k}_{2})\cdot{\bf r}}\
     F_{\pi}({\bf k}_{1}^{2}) F_{\eta}({\bf k}_{2}^{2})      \nonumber\\
   &\times& ({\bf k}_{1}\!\cdot\!{\bf k}_{2})
     \biggl[ ({\bf k}_{1}\!\cdot\!{\bf k}_{2})^{2}
     -\bbox{\sigma}_{1}\!\cdot\!({\bf k}_{1}\times{\bf k}_{2})
      \bbox{\sigma}_{2}\!\cdot\!({\bf k}_{1}\times{\bf k}_{2})
\nonumber\\ && \hspace{1.2cm}
     +i(\bbox{\sigma}_{1}+\bbox{\sigma}_{2})\!\cdot\!
       ({\bf k}_{1}\times{\bf k}_{2})
       {\bf q}\!\cdot\!({\bf k}_{1}-{\bf k}_{2}) \biggr]
     D_{//}^{(1)}(\omega_{1},\omega_{2}).               \label{nonadPVp}
\end{eqnarray}
\narrowtext

All integrals in Eqs.~(\ref{nonadS})--(\ref{nonadPVp}) can be
readily evaluated using the results of Appendix~\ref{app:appA}.
Inspection of the energy denominators
$D_{i}^{(1)}(\omega_{1},\omega_{2})$ reveals that we need derivatives
of the function $I_{4}(m,r)$, which is defined by~\cite{Rij91}
\begin{equation}
    I_{4}(m,r)=-\frac{d}{dm^{2}}I_{2}(m,r)
               +\frac{1}{\Lambda^{2}}I_{2}(m,r),     \label{I4fun}
\end{equation}
while $I_{4}(m_{P},r)$ for the Pomeron is given by the second
expression in Eq.~(\ref{pom7}).

\subsection{Pseudovector-vertex corrections}
\label{sec:chap5b}
The pseudovector vertex gives rise to $1/M$ terms as shown in
Eq.~(\ref{NNPV}). In the planar-box diagrams, the $1/M$ contribution
from each diagram is cancelled by the $1/M$ contribution from its
``mirror'' diagram, and therefore we only have to consider the
crossed-box BW diagrams. The pseudovector-vertex corrections are
very simple to derive for the pion-scalar and pion-vector potentials.
Below, we again take a representative meson for each type. Because of
their simplicity, we also give the explicit coordinate-space
expressions for these potentials.
For pion-scalar exchange we get
\begin{eqnarray}
  V^{(1b)}_{\pi\varepsilon} &=& C_{N\!N}^{(1,0)}(I)
    g_{N\!N\!\varepsilon}^{2}
    \left(\frac{f_{N\!N\!\pi}}{m_{\pi}}\right)^{2}
    \left(\frac{1}{M}\right)                                 \nonumber\\
  &&\times \int\!\!\int\frac{d^{3}k_{1}d^{3}k_{2}}{(2\pi)^{6}}\,
    e^{i({\bf k}_{1}+{\bf k}_{2})\cdot{\bf r}}\
    F_{\pi}({\bf k}_{1}^{2})F_{\varepsilon}({\bf k}_{2}^{2}) \nonumber\\
  &&\times \biggl[(\bbox{\sigma}_{1}\!\cdot\!{\bf k}_{1})
                  (\bbox{\sigma}_{2}\!\cdot\!{\bf k}_{2}) +
                  (\bbox{\sigma}_{1}\!\cdot\!{\bf k}_{2})
                  (\bbox{\sigma}_{2}\!\cdot\!{\bf k}_{1}) \biggr]
      \frac{1}{2\omega_{1}^{2}\omega_{2}^{2}}                \nonumber\\
  &=& -C_{N\!N}^{(1,0)}(I) g_{N\!N\!\varepsilon}^{2}
    \left(\frac{f_{N\!N\!\pi}}{m_{\pi}}\right)^{2}
    \left(\frac{1}{3M}\right)                                \nonumber\\
  &&\times I'_{2}(m_{\pi},r)I'_{2}(m_{\varepsilon},r)\,
    [(\bbox{\sigma}_{1}\!\cdot\!\bbox{\sigma}_{2})+S_{12}].\label{pvappS}
\end{eqnarray}
For pion-vector exchange we get
\begin{eqnarray}
  V^{(1b)}_{\pi\!\rho} &=& -C_{N\!N}^{(X)}(I)
    g_{N\!N\!\rho}^{2}\left(\frac{f_{N\!N\!\pi}}{m_{\pi}}\right)^{2}
    \left(\frac{1}{M}\right)                                 \nonumber\\
  &&\times \int\!\!\int\frac{d^{3}k_{1}d^{3}k_{2}}{(2\pi)^{6}}\,
    e^{i({\bf k}_{1}+{\bf k}_{2})\cdot{\bf r}}\
    F_{\pi}({\bf k}_{1}^{2}) F_{\rho}({\bf k}_{2}^{2})       \nonumber\\
  &&\times \biggl[(\bbox{\sigma}_{1}\!\cdot\!{\bf k}_{1})
                  (\bbox{\sigma}_{2}\!\cdot\!{\bf k}_{2}) +
                  (\bbox{\sigma}_{1}\!\cdot\!{\bf k}_{2})
                  (\bbox{\sigma}_{2}\!\cdot\!{\bf k}_{1}) \biggr]
      \frac{1}{2\omega_{1}^{2}\omega_{2}^{2}}                \nonumber\\
  &=& C_{N\!N}^{(X)}(I)g_{N\!N\!\rho}^{2}
    \left(\frac{f_{N\!N\!\pi}}{m_{\pi}}\right)^{2}
    \left(\frac{1}{3M}\right)                                \nonumber\\
  &&\times I'_{2}(m_{\pi},r)I'_{2}(m_{\rho},r)\,
    [(\bbox{\sigma}_{1}\!\cdot\!\bbox{\sigma}_{2})+S_{12}]. \label{pvappV}
\end{eqnarray}

The pion-pseudoscalar case is more complicated than the pion-scalar
and pion-vector cases, because now we also have to include the $1/M$
corrections from the pseudovector vertex of the $\eta$ or $\eta'$ meson.
Again, the contributions from the planar-box diagrams are cancelled
by their ``mirror'' diagrams. Including the ``time-reversed'' diagrams,
and defining the function
\begin{equation}
    I_{0}(\Lambda,r)=\frac{1}{4\pi}\frac{\Lambda^{3}}{2\sqrt{\pi}}
             e^{-\frac{1}{4}\Lambda^{2}r^{2}},
\end{equation}
we find for the crossed diagrams
\widetext
\begin{eqnarray}
  V^{(1b)}_{\pi\eta} &=& C_{N\!N}^{(1,0)}(I)
    \left(\frac{f_{N\!N\!\eta}}{m_{\pi}}\right)^{2}
    \left(\frac{f_{N\!N\!\pi}}{m_{\pi}}\right)^{2}
    \left(\frac{1}{M}\right)
          \int\!\!\int\frac{d^{3}k_{1}d^{3}k_{2}}{(2\pi)^{6}}\,
    e^{i({\bf k}_{1}+{\bf k}_{2})\cdot{\bf r}}\
    F_{\pi}({\bf k}_{1}^{2}) F_{\eta}({\bf k}_{2}^{2})       \nonumber\\
  &&\times \biggl\{ ({\bf k}_{1}\!\cdot\!{\bf k}_{2})
        ({\bf k}_{1}^{2}+{\bf k}_{2}^{2})
% \nonumber\\ && \ \ \
       +i(\bbox{\sigma}_{1}+\bbox{\sigma}_{2})\!\cdot\!\biggl[
         {\bf q}\times({\bf k}_{1}+{\bf k}_{2})
       ({\bf k}_{1}\!\cdot\!{\bf k}_{2}) + ({\bf k}_{1}\times{\bf k}_{2})\,
         {\bf q}\!\cdot\!({\bf k}_{1}-{\bf k}_{2}) \biggr] \biggr\}
       \frac{1}{\omega_{1}^{2}\omega_{2}^{2}}                \nonumber\\
  &=& C_{N\!N}^{(1,0)}(I)\left(\frac{f_{N\!N\!\eta}}{m_{\pi}}\right)^{2}
    \left(\frac{f_{N\!N\!\pi}}{m_{\pi}}\right)^{2}
    \left(\frac{1}{M}\right)                                 \nonumber\\
  &&\times \biggl\{  \biggl[ \left(m_{\pi}^{2}+m_{\eta}^{2}\right)
        I'_{2}(m_{\pi},r)I'_{2}(m_{\eta},r)
       -I'_{0}(\Lambda_{\pi},r)I'_{2}(m_{\eta},r)
       -I'_{2}(m_{\pi},r)I'_{0}(\Lambda_{\eta},r) \biggr]    \nonumber\\
  &&\ \ +\left[\frac{2}{r}\biggl(I''_{2}(m_{\pi},r)I'_{2}(m_{\eta},r)
            +I'_{2}(m_{\pi},r)I''_{2}(m_{\eta},r)\biggr)
       +\frac{4}{r^{2}}I'_{2}(m_{\pi},r)I'_{2}(m_{\eta},r)\right]
        {\bf L}\!\cdot\!{\bf S} \biggr\}.                \label{pvappPV}
\end{eqnarray}
\narrowtext

\subsection{Off-shell corrections in TMO diagrams}
\label{sec:chap5c}
Next to the type of nonadiabatic (off-shell) corrections discussed in
Sec.~\ref{sec:chap5a}, there are $1/M$ corrections in the TMO diagrams,
due to off-shell factors like $({\bf p}^{\prime\prime2}-{\bf p}^{2})$
from the pseudovector vertices [see Eq.~(\ref{NNPV})].
In the on-shell approximation, ${\bf p}^{\prime2}-{\bf p}_{i}^{2}=
{\bf p}^{2}-{\bf p}_{i}^{2}=0$, this type of momentum dependence in
the numerator cancels the ${\bf p}^{\prime\prime2}-{\bf p}_{i}^{2}$
momentum dependence of the two-nucleon intermediate state in the energy
denominator of the TMO diagrams, leaving a $1/M$ correction to the
potential.

For the pion-scalar and pion-vector exchanges it can be easily
shown that these (what we call) off-shell TMO contributions
involve the momentum operator
\begin{equation}
   O_{\rm off}({\bf k}_{1},{\bf k}_{2})=
   \frac{1}{M}\left[\bbox{\sigma}_{1}\!\cdot\!{\bf k}_{1}
                    \bbox{\sigma}_{2}\!\cdot\!{\bf k}_{1}
        \mp{\textstyle\frac{1}{2}}
           (\bbox{\sigma}_{1}\!\cdot\!{\bf k}_{1}
            \bbox{\sigma}_{2}\!\cdot\!{\bf k}_{2}
           +\bbox{\sigma}_{1}\!\cdot\!{\bf k}_{2}
            \bbox{\sigma}_{2}\!\cdot\!{\bf k}_{1}\right], \label{Off}
\end{equation}
where the $\mp$ sign refers to pseudovector or pseudoscalar
coupling for the pion, respectively. Hence
\begin{eqnarray}
  V^{(1c)}_{\pi\varepsilon} &=& -C_{N\!N}^{(1,0)}(I)
    g_{N\!N\!\varepsilon}^{2}
    \left(\frac{f_{N\!N\!\pi}}{m_{\pi}}\right)^{2}
    \left(\frac{1}{2M}\right)                                \nonumber\\
  &&\times F_{\pi}({\bf k}_{1}^{2})F_{\varepsilon}({\bf k}_{2}^{2})
    O_{\rm off}({\bf k}_{1},{\bf k}_{2})
    \frac{1}{\omega_{1}^{2}\omega_{2}^{2}},               \label{offS}\\
  V^{(1c)}_{\pi\rho} &=& C_{N\!N}^{(//)}(I)
    g_{N\!N\!\rho}^{2}\left(\frac{f_{N\!N\!\pi}}{m_{\pi}}\right)^{2}
    \left(\frac{1}{2M}\right)                                \nonumber\\
  &&\times F_{\pi}({\bf k}_{1}^{2})F_{\rho}({\bf k}_{2}^{2})
    O_{\rm off}({\bf k}_{1},{\bf k}_{2})
    \frac{1}{\omega_{1}^{2}\omega_{2}^{2}}.                \label{offV}
\end{eqnarray}
Because the off-shell factors in the pseudovector vertices occur with
the same ($\bbox{\sigma}_{1},\bbox{\sigma}_{2}$) spin dependence as
the $\omega/M$ terms, it should be obvious that the resulting potentials
are very similar to the pseudovector-vertex corrections derived in the
previous section (except for a relative minus sign in the spin-orbit
pion-pseudoscalar part). We find
\begin{eqnarray}
  V^{(1c)}_{\pi\eta} &=& C_{N\!N}^{(1,0)}(I)
    \left(\frac{f_{N\!N\!\eta}}{m_{\pi}}\right)^{2}
    \left(\frac{f_{N\!N\!\pi}}{m_{\pi}}\right)^{2}
    \left(\frac{1}{2M}\right)                                \nonumber\\
  &&\times \biggl\{  \biggl[ \left(m_{\pi}^{2}+m_{\eta}^{2}\right)
        I'_{2,\pi}I'_{2,\eta} -I'_{0,\pi}I'_{2,\eta}
       -I'_{2,\pi}I'_{0,\eta} \biggr]    \nonumber\\
  &&\ \ +\left[\frac{2}{r}
         \biggl(I''_{2,\pi}I'_{2,\eta}+I'_{2,\pi}I''_{2,\eta}\biggr)
       -\frac{4}{r^{2}}I'_{2,\pi}I'_{2,\eta}\right]
        {\bf L}\!\cdot\!{\bf S} \biggr\}.                    \nonumber\\
  && \label{offPV}
\end{eqnarray}

\section{RESULTS AND DISCUSSION}
\label{sec:chap7}
The complete pion-meson-exchange potential can be written as a sum of
all pion-meson exchanges $V(\pi\alpha)$, where
\begin{equation}
 V(\pi\alpha)=\sum_{n=0,1a,1b,1c} \left[ V^{(n)}_{\pi\alpha}(//)+
                                 V^{(n)}_{\pi\alpha}(X)\right].
\end{equation}
Here, the second meson is denoted by $\alpha$. Each potential consists
of central, spin-spin, tensor, and spin-orbit parts.
The leading-order contributions to the coordinate-space pion-meson
potentials are given explicitly in Appendix~\ref{app:appB}.
The nonadiabatic $1/M$ corrections can be easily derived
from Sec.~\ref{sec:chap5a} and Appendix~\ref{app:appA},
the pseudovector $1/M$ corrections are given in Sec.~\ref{sec:chap5b},
and the extra $1/M$ corrections in the TMO diagrams are given in
Sec.~\ref{sec:chap5c}.

In the following we show the various contributions to the potentials,
using the meson-nucleon coupling constants and cutoff masses as given
in Table~\ref{table4}. They correspond to the values of a preliminary
version of the Nijmegen extended soft-core (ESC) potential which,
next to the standard one-boson exchanges and the two-meson exchanges
discussed in this paper, also contains the pair-meson exchanges to be
discussed in the following paper~\cite{Rij95b}.
In this model, the pion-nucleon coupling constant at the pion pole
was fixed at $f_{N\!N\pi}^{2}/4\pi=0.0745$, in accordance with the
recommended value~\cite{Sto93a}. Keeping in line with the philosophy
of the Nijmegen group~\cite{Nag78,Mae89,Sto94}, we use SU(3) relations
for the coupling constants. However, since the existence of a scalar
nonet (and hence its quark content) is still controversial, the three
nucleon-nucleon-scalar coupling constants are fitted independently.
To further limit the number of free parameters, the value for
$(f/g)_{\rho}$ was fixed at the vector-meson dominance value of 3.71.
The result of $(f/g)_{\omega}=-0.06$ is also found to be very close to
the vector-meson dominance value of --0.12, but is not expected to
hold exactly since the $\phi$ meson is not considered to be a pure
$s\bar{s}$ system (we do not assume ideal mixing for the vector mesons).
It is also interesting to note that the $\varepsilon$ coupling constant
is much smaller than in the one-boson-exchange Nijm93
potential~\cite{Sto94}, which is to be expected since in the present
ESC model two-pion exchanges are explicitly included.
The 14 free parameters are fitted to the 1993 Nijmegen representation
of the $\chi^{2}$ hypersurface of the $N\!N$ scattering data below
$T_{\rm lab}=350$ MeV~\cite{Sto93b}, updated with the inclusion of
new data which have been published since then.
A comparison between this updated partial-wave analysis and the Nijm93
and ESC potentials will be given in the companion paper~\cite{Rij95b}.

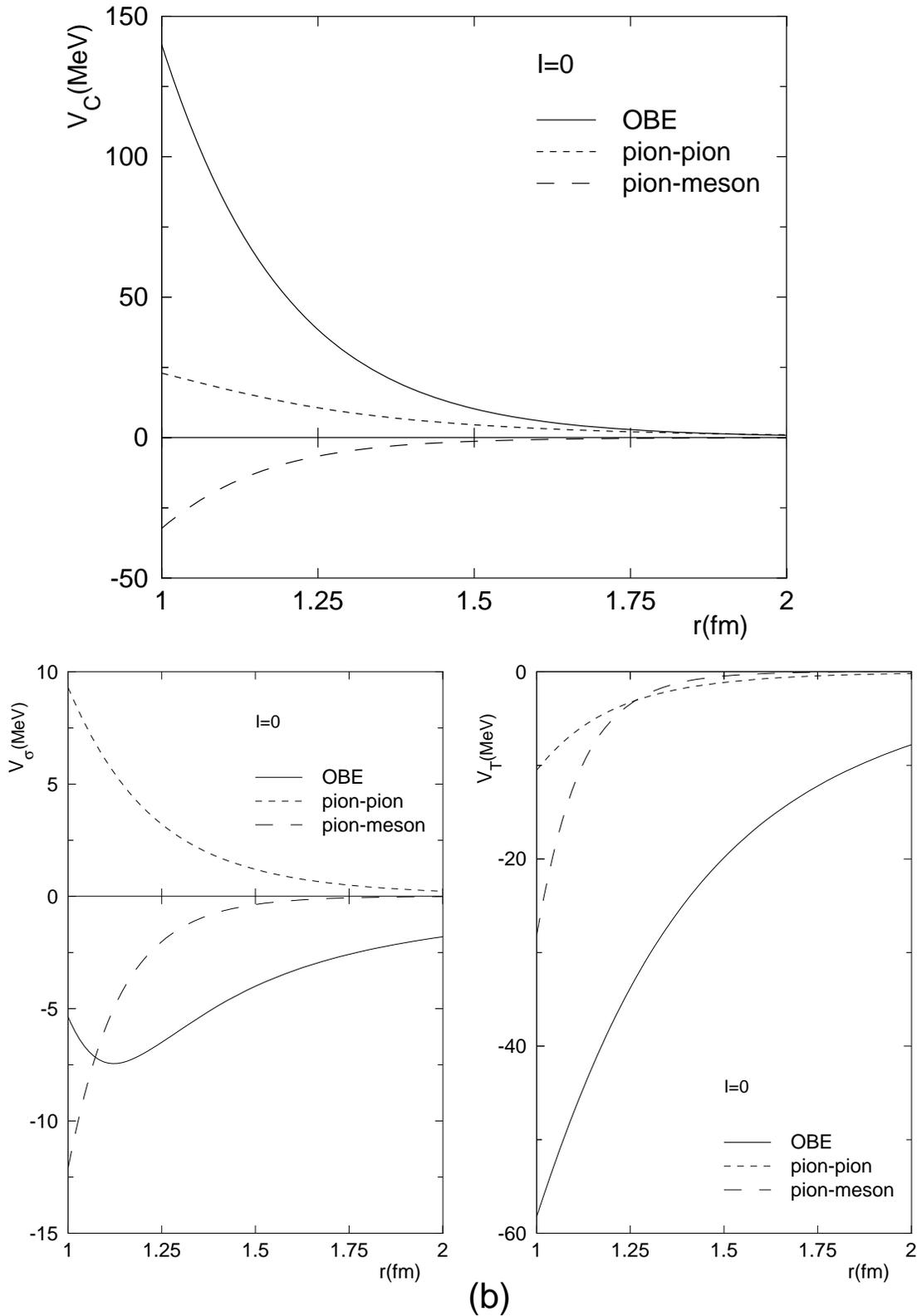
\begin{figure}
\caption{Central, spin-spin, and tensor potentials for $I=0$.
         Shown are the one-boson exchange (OBE), the pion-pion exchange,
         and the other pion-meson exchanges.
         All potentials are for (a) $r\leq1$ fm and (b) $1\leq r\leq2$
         fm. The spin-spin pion-meson potential has its maximum at
         4874 MeV.}
\label{pap1fig3}
\end{figure}

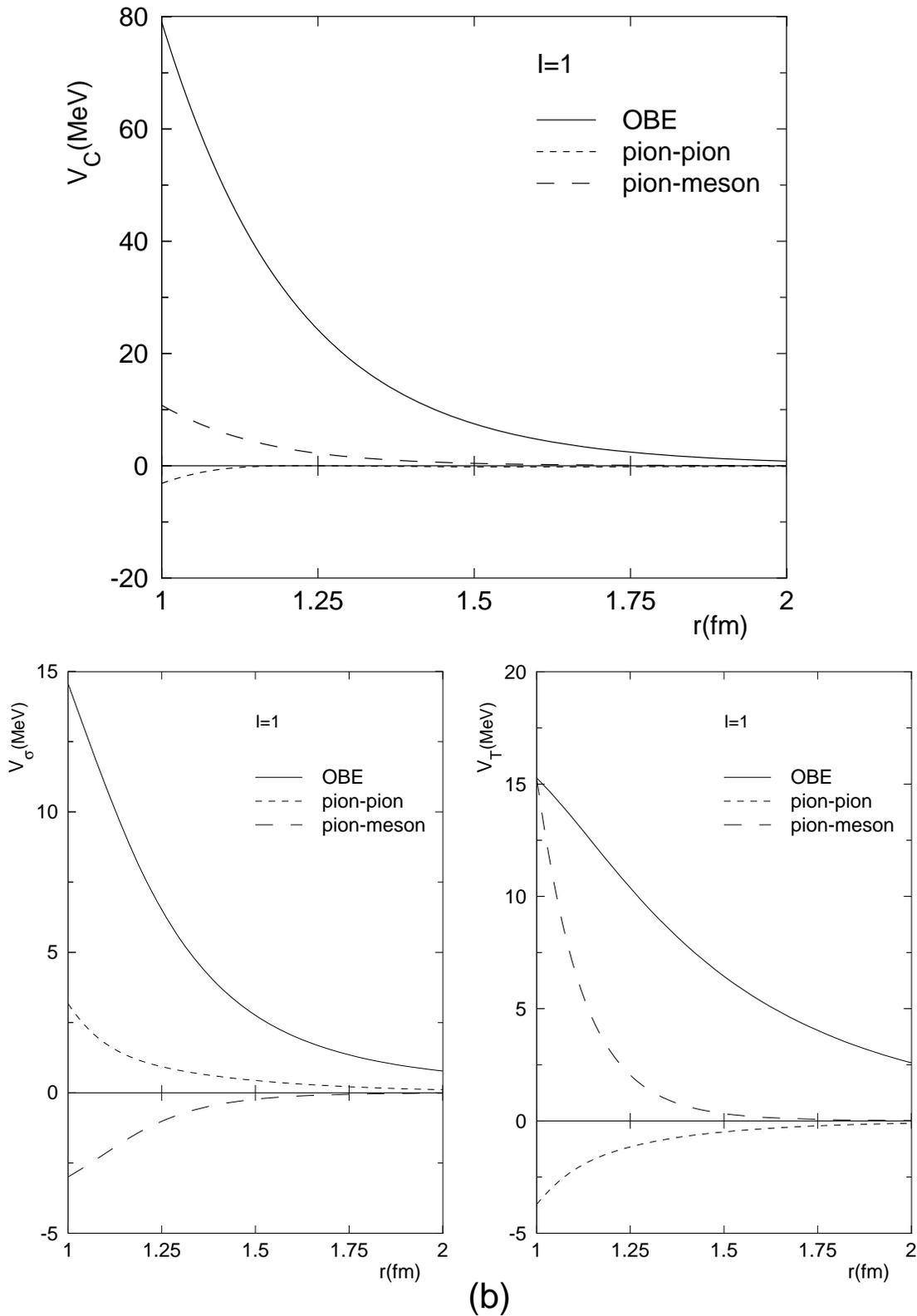
\begin{figure}
\caption{Same as Fig.~\protect\ref{pap1fig3}, but for $I=1$. The
         spin-spin pion-meson potential has its minimum at --2969 MeV.}
\label{pap1fig4}
\end{figure}

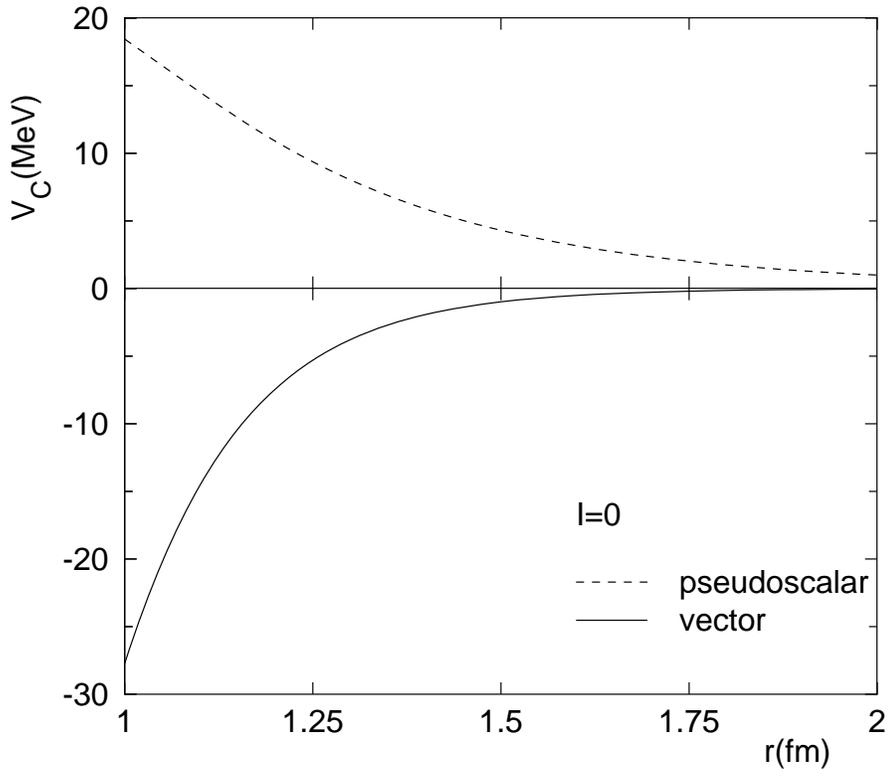
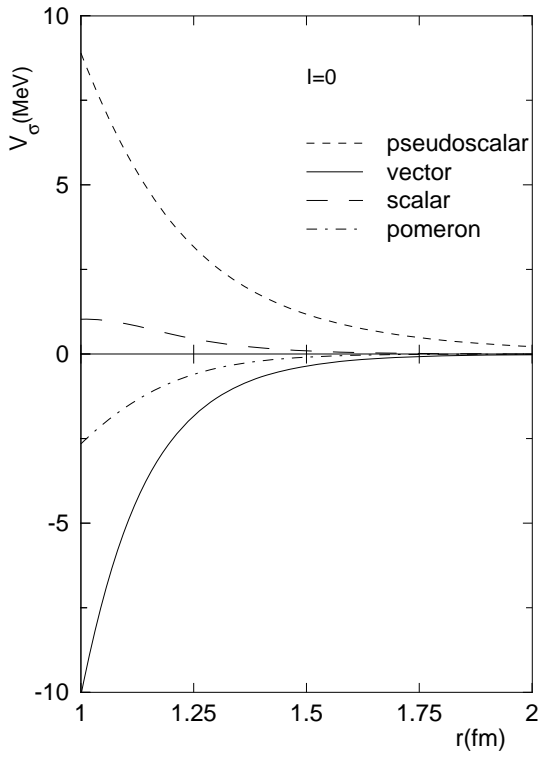
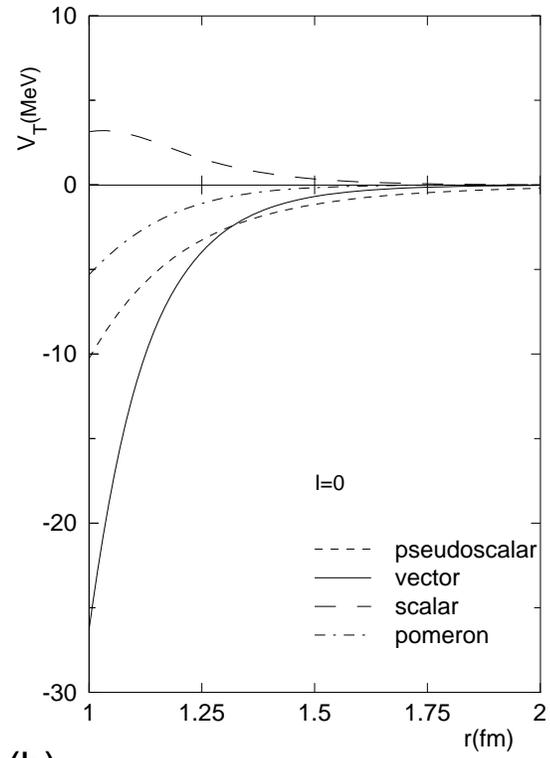
\begin{figure}
\caption{Central, spin-spin, and tensor potentials for $I=0$.
         Shown are the pion-pseudoscalar, pion-vector, pion-scalar,
         and pion-Pomeron exchanges.
         All potentials are for (a) $r\leq1$ fm and (b) $1\leq r\leq2$
         fm.}
\label{pap1fig5}
\end{figure}

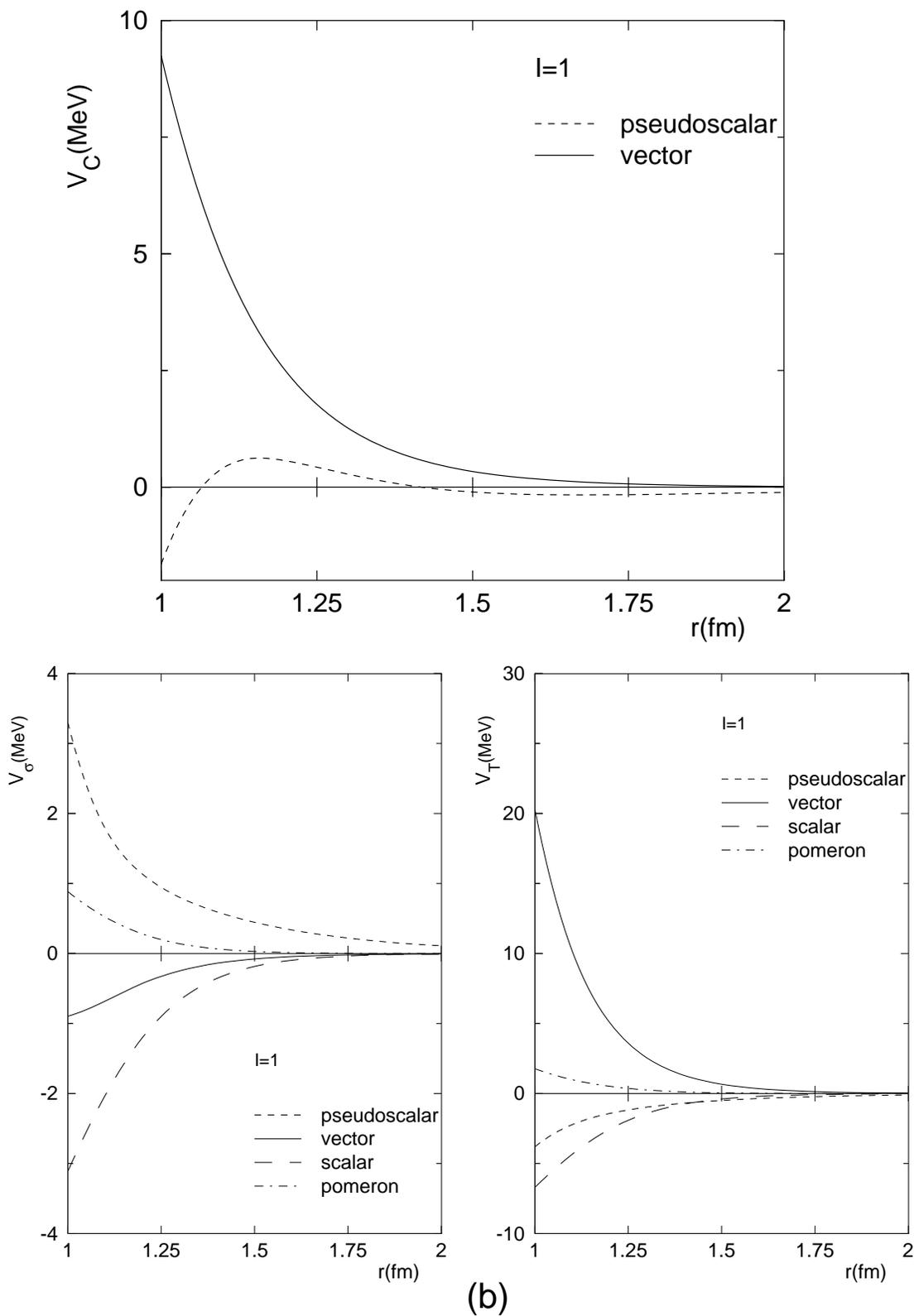
\begin{figure}
\caption{Same as Fig.~\protect\ref{pap1fig5}, but for $I=1$.}
\label{pap1fig6}
\end{figure}

In Figs.~\ref{pap1fig3} and \ref{pap1fig4}, we compare the contributions
from one-boson exchange, pion-pion exchange, and the other pion-meson
exchanges; all for isospin $I=0$ and $I=1$, respectively.
The OBE contributions are the standard OBE potentials, substituting
the coupling constants and cutoff masses as given in Table~\ref{table4}.
Clearly, these OBE contributions are different from what one would
obtain in a potential model containing {\it only\/} OBE contributions:
the presence of the two-meson contributions substantially modifies
the short-range behavior which has to be compensated for. Hence, the
OBE contributions of the present ESC model on their own will certainly
not fit the data.
All potentials are seen to level off towards moderate values at the
origin. This is due to the Gaussian form factors.
(Note that the Gaussian form factors ensure that all potentials are
finite at the origin to all orders in ${\bf k}^{2}/M^{2}$ and that
the tensor components always vanish at the origin; this in contrast
to a monopole or dipole form factor.)
The pion-meson contributions dominate the inner region; especially the
spin-spin potential, which is mainly due to pion-vector exchange
(see Figs.~\ref{pap1fig5} and \ref{pap1fig6}).
For $I=1$ the one-boson and pion-pion central contributions cancel
each other to a large extent, whereas for $I=0$ they enhance each other.
The OBE contributions clearly dominate the outer region ($r\gtrsim1$ fm).

In Figs.~\ref{pap1fig5} and \ref{pap1fig6}, we compare the contributions
to the pion-meson potentials from the different types of mesons:
pseudoscalar ($\pi$, $\eta$, $\eta'$), vector ($\rho$, $\omega$, $\phi$),
scalar ($a_{0}$, $\varepsilon$, $f_{0}$), and Pomeron.
The pion-scalar and pion-Pomeron potentials do not have a central
component, while the central component of the $\pi\eta$ and $\pi\eta'$
potentials is very small, as can be inferred from the comparison of the
pseudoscalar curve in Figs.~\ref{pap1fig5} and \ref{pap1fig6} and the
pion-pion curve in Figs.~\ref{pap1fig3} and \ref{pap1fig4}, respectively.
A similar comparison for the spin-spin and tensor components shows that,
at short distances, the $\pi\eta$ and $\pi\eta'$ potentials oppose the
$\pi\pi$ potentials for $I=0$, resulting in a largely reduced
pion-pseudoscalar contribution. For $I=1$ they slightly enhance the
$\pi\pi$ potentials.
Looking into even more detail (not shown in the figures), we find
that the $\pi\eta$ and $\pi\eta'$ potentials are the least important.
This is due to the small $N\!N\eta$ and $N\!N\eta'$ coupling constants
and the high mass of the $\eta'$.

The large spin-spin pion-vector potential can be traced to the rather
large $\omega$ coupling constant and cutoff mass.
At the origin, the $\pi\omega$ component reaches a maximum of 3122 MeV
for $I=0$, which is only partially reduced by the $\pi\rho$ component.
For $I=1$ the $\pi\omega$ component reaches a minimum of --1040 MeV,
which enhances the equally large $\pi\rho$ component. The $\pi\phi$
component contributes much less, as is to be expected due to the large
$\phi$-meson mass. At present, it is not yet clear to us why the
model apparently requires such a large spin-spin component.

Comparing our results to those of Holinde and Machleidt (HM)~\cite{Hol81}
we also find a large cancelation between the $\pi\pi$ and $\pi\rho$
potentials. The cancelation between the $\pi\varepsilon$ and $\pi\omega$
potentials, however, is much less pronounced. One reason for this is
the enormous difference in coupling constants: we have
$g_{\varepsilon}^{2}/4\pi=0.34$ and $g_{\omega}^{2}/4\pi=11$, whereas
HM use 6 and 23, respectively. As a matter of fact, in our model the
contribution of the $\pi\varepsilon$ potential is almost negligible.
Another reason for the only partial cancelation is that we
include the scalar $f_{0}$ and vector $\phi$ mesons, whereas HM do not.

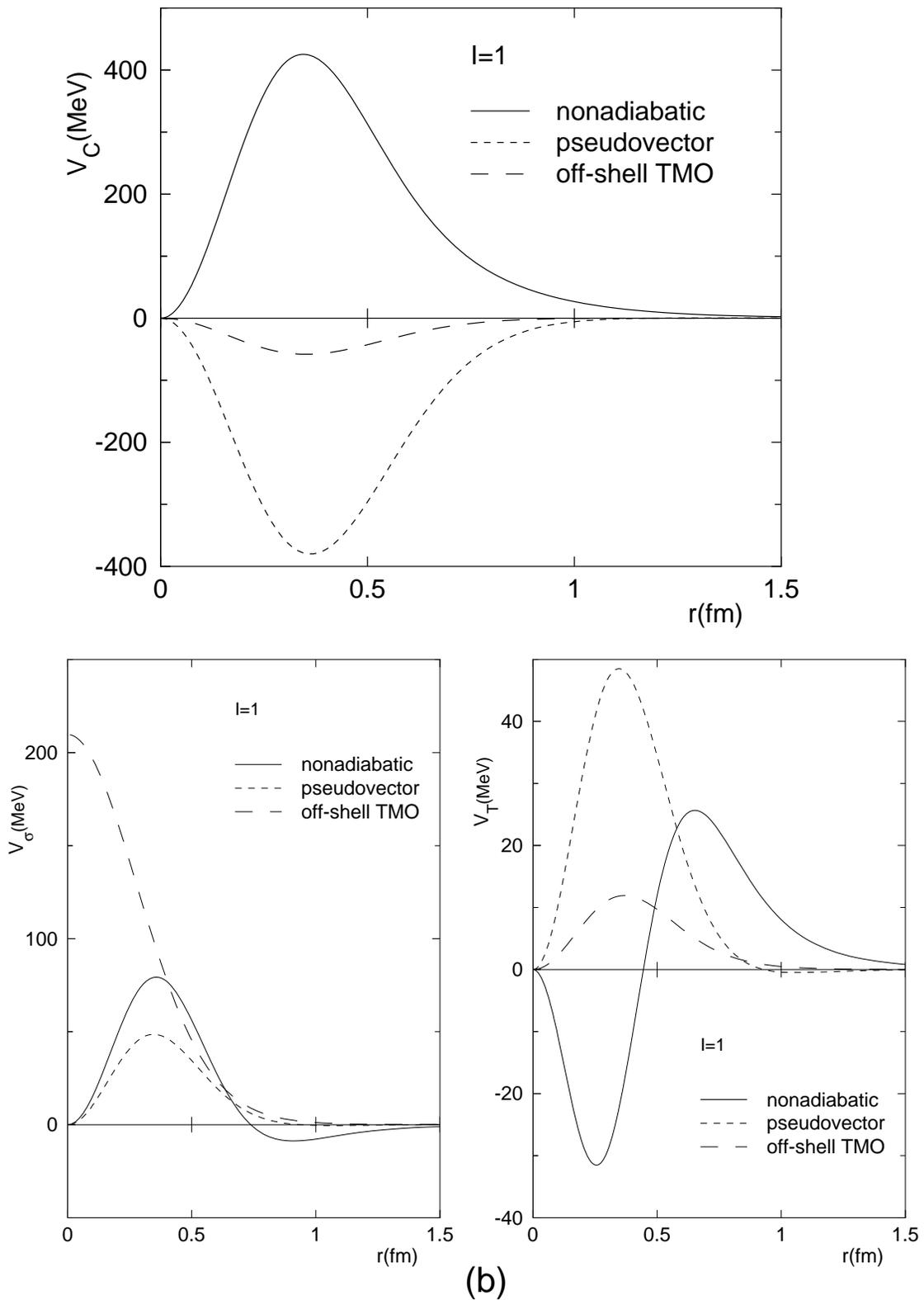
\begin{figure}
\caption{Nonadiabatic, pseudovector-vertex, and purely off-shell
         $1/M$ contributions to the pion-meson potentials for (a)
         $I=0$ and (b) $I=1$.}
\label{pap1fig7}
\end{figure}

The $1/M$ contributions to the potential due to the nonadiabatic,
pseudovector-vertex, and off-shell TMO $1/M$ corrections are shown in
Fig.~\ref{pap1fig7}. Except for the off-shell TMO spin-spin part, all
contributions are seen to vanish at the origin. Furthermore, there are
large cancelations between the nonadiabatic and pseudovector-vertex
corrections, and so the combined $1/M$ contribution to the pion-meson
potential is indeed much smaller than the leading-order contribution,
validating our procedure of expanding the potential as a series in $1/M$.

In Fig.~\ref{pap1fig8} we show the spin-orbit contributions of the
pion-meson potentials in comparison with the spin-orbit OBE potential.
The spin-orbit potentials of the pion-scalar and pion-vector exchanges
are due to the $1/M^{2}$ terms in Eqs.~(\ref{pots}) and (\ref{potv}).
The spin-orbit pion-pseudoscalar potential, on the other hand, is
mainly due to the $1/M$ pseudovector-vertex contributions, and hence
is much larger.
For $I=1$, the pion-pseudoscalar spin-orbit potential largely
cancels the OBE spin-orbit potential in the inner region, whereas
for $I=0$ they enhance each other. The resulting spin-orbit part
is reduced by the pion-vector spin-orbit contribution. (Incidentally,
the large $\omega$ coupling constant and cutoff mass might be
a necessary condition to arrive at a sufficiently moderate $I=0$
spin-orbit potential.)
On the other hand, the $1/M$ nonadiabatic and off-shell TMO spin-orbit
contributions, though much smaller than the pseudovector-vertex
contribution, peak near 0.4 fm and, therefore, cause a substantially
slower fall-off of the spin-orbit component. Because the spin-orbit
operator ${\bf L}\!\cdot\!{\bf S}$ is proportional to the angular
momentum $L$, this means that this contribution becomes more and more
important for the higher partial waves, i.e., at higher energies.
We indeed find that the quality of the fit rapidly becomes worse
beyond $\sim$300 MeV. It has been argued~\cite{Kle53,Lev52} that these
spin-orbit contributions will be (partially) cancelled by the inclusion
of higher-order (three-meson- and four-meson-exchange) diagrams, and so
perhaps they should not be included at the present level of our
expansion. Alternatively, one can argue that the nonadiabatic
expansion in principle breaks down at the pion-production threshold
($\sim$280 MeV), and so the energy range should not be extended to
350 MeV anyway. However, the inclusion of the nonadiabatic central,
spin-spin, and tensor contributions does provide a considerable
improvement beyond $\sim$280 MeV, apparently contradicting this
line of argument. Currently, we are still investigating this matter.

Finally, we note that if we choose to take the pseudoscalar coupling
(\ref{Lagps}) instead of the pseudovector coupling (\ref{Lagpv}), there
are some small differences in the pion-scalar and pion-vector potentials;
i.e., one of the $1/M^{2}$ terms changes sign, see Table~\ref{table2}.
A much larger effect is due to the fact that in case of the
pseudoscalar coupling the $\omega/M$ pseudovector-vertex corrections
of Sec.~\ref{sec:chap5b} are all absent. Also, there is a sign change in
the off-shell TMO corrections, which means that in the pion-pseudoscalar
potential the nonadiabatic and off-shell TMO spin-orbit contributions
exactly cancel (they have the same $r$ dependence). Hence, the
apparent problem with the spin-orbit contribution as discussed above
does not occur.

\acknowledgments
We would like to thank Prof.\ J.J.\ de Swart, Prof.\ I.R.\ Afnan, and
the other members of the theory groups at Nijmegen and Flinders for
their stimulating interest.
The work of V.S.\ was financially supported by the Australian Research
Council.

\widetext
\appendix
\section{DIFFERENTIATION DICTIONARY}
\label{app:appA}
In this Appendix we give a dictionary for the evaluation of the
differentiations in Eqs.~(\ref{potps}), (\ref{pots}),
(\ref{potv}), and the expressions in Sec.~\ref{sec:chap5}.
The procedure is described in more detail in Appendix B of \cite{Rij92a}.
In the following it is to be understood that the functions on the
right-hand side are functions of $r$ and that the prime denotes
differentiation with respect to $r$.

\noindent 1. For the pseudoscalar operators $O^{(i)}_{PS}$:
\begin{eqnarray*}
 & & \lim_{{\bf r}_{1}\rightarrow{\bf r}_{2}}
   (\bbox{\nabla}_{1}\!\cdot\!\bbox{\nabla}_{2})^{2}F(r_{1})G(r_{2}) =
    \frac{2}{r^{2}}F'G'+F''G'',                                        \\
 & & \lim_{{\bf r}_{1}\rightarrow{\bf r}_{2}}
   (\bbox{\sigma}_{1}\!\cdot\!\bbox{\nabla}_{1}\!\times\!\bbox{\nabla}_{2})
   (\bbox{\sigma}_{2}\!\cdot\!\bbox{\nabla}_{1}\!\times\!\bbox{\nabla}_{2})
        F(r_{1})G(r_{2}) = \frac{2}{3}\left[\frac{1}{r^{2}}F'G'
         +\frac{1}{r}F'G''+\frac{1}{r}F''G'\right]
         (\bbox{\sigma}_{1}\!\cdot\!\bbox{\sigma}_{2})       \nonumber\\
 & & \hspace{7.0cm} -\frac{1}{3}\left[\left(F''-\frac{1}{r}F'\right)
        \frac{1}{r}G'+\frac{1}{r}F'\left(G''-\frac{1}{r}G'\right)
     \right] S_{12}.
\end{eqnarray*}

\noindent 2. For the scalar operators $O^{(i)}_{S}$ and the vector
operators $O^{(i)}_{e,e}$, $O^{(i)}_{e,m}$, $O^{(i)}_{e,2m}$,
and $O^{(i)}_{m,m}$:
\begin{eqnarray*}
 & & \lim_{{\bf r}_{1}\rightarrow{\bf r}_{2}}
   (\bbox{\sigma}_{1}\!\cdot\!\bbox{\nabla}_{1})
   (\bbox{\sigma}_{2}\!\cdot\!\bbox{\nabla}_{1})F(r_{1})G(r_{2}) =
   \frac{1}{3}\left[\left(F''+\frac{2}{r}F'\right)G
         (\bbox{\sigma}_{1}\!\cdot\!\bbox{\sigma}_{2})
                  + \left(F''-\frac{1}{r}F'\right)G\, S_{12}\right], \\
 & & \lim_{{\bf r}_{1}\rightarrow{\bf r}_{2}}
   (\bbox{\sigma}_{1}\!\cdot\!\bbox{\nabla}_{1})
   (\bbox{\sigma}_{2}\!\cdot\!\bbox{\nabla}_{1})
    \bbox{\nabla}_{2}^{2}F(r_{1})G(r_{2}) = \frac{1}{3}\left[
      \left(F''+\frac{2}{r}F'\right)\left(G''+\frac{2}{r}G'\right)
         (\bbox{\sigma}_{1}\!\cdot\!\bbox{\sigma}_{2})
\right. \\ & & \hspace{7.5cm} \left.
     +\left(F''-\frac{1}{r}F'\right)
         \left(G''+\frac{2}{r}G'\right) S_{12}\right],            \\
 & & \lim_{{\bf r}_{1}\rightarrow{\bf r}_{2}}
    (\bbox{\sigma}_{1}\!\cdot\!\bbox{\nabla}_{1})
    (\bbox{\sigma}_{2}\!\cdot\!\bbox{\nabla}_{1})
    (\bbox{\nabla}_{1}\!\cdot\!\bbox{\nabla}_{2})F(r_{1})G(r_{2}) =
    \frac{1}{3}\left[\left(F'''+\frac{2}{r}F''-\frac{2}{r^{2}}F'\right)G'
    (\bbox{\sigma}_{1}\!\cdot\!\bbox{\sigma}_{2})
\right. \\ & & \hspace{7.5cm} \left.
    +\left(F'''-\frac{1}{r}F''+\frac{1}{r^{2}}F'\right)G'\,
     S_{12}\right],                                   \\
 & & \lim_{{\bf r}_{1}\rightarrow{\bf r}_{2}}
   (\bbox{\sigma}_{1}\!\cdot\!\bbox{\nabla}_{1})
   (\bbox{\sigma}_{2}\!\cdot\!\bbox{\nabla}_{2})
    \bbox{\nabla}_{1}^{2}F(r_{1})G(r_{2}) = \frac{1}{3}
    \left[F'''+\frac{2}{r}F''-\frac{2}{r^{2}}F'\right] G'
    \left[(\bbox{\sigma}_{1}\!\cdot\!\bbox{\sigma}_{2})+S_{12}\right], \\
 & & \lim_{{\bf r}_{1}\rightarrow{\bf r}_{2}}
   (\bbox{\sigma}_{1}\!\cdot\!\bbox{\nabla}_{2})
   (\bbox{\sigma}_{2}\!\cdot\!\bbox{\nabla}_{1})
    \bbox{\nabla}_{1}^{2}F(r_{1})G(r_{2}) = \frac{1}{3}
    \left[F'''+\frac{2}{r}F''-\frac{2}{r^{2}}F'\right] G'
    \left[(\bbox{\sigma}_{1}\!\cdot\!\bbox{\sigma}_{2})+S_{12}\right], \\
 & & \lim_{{\bf r}_{1}\rightarrow{\bf r}_{2}}
   (\bbox{\sigma}_{1}\!\cdot\!\bbox{\nabla}_{1})
   (\bbox{\sigma}_{2}\!\cdot\!\bbox{\nabla}_{1})
    \bbox{\nabla}_{1}^{2}F(r_{1})G(r_{2}) = \frac{1}{3}\left[
    \left(F''''+\frac{4}{r}F'''\right)G
        (\bbox{\sigma}_{1}\!\cdot\!\bbox{\sigma}_{2})
\right. \\ & & \hspace{7.5cm} \left.
   +\left(F''''+\frac{1}{r}F'''-\frac{6}{r^{2}}F''
    +\frac{6}{r^{3}}F'\right)G\, S_{12}\right],                      \\
 & & \lim_{{\bf r}_{1}\rightarrow{\bf r}_{2}}
    [{\bf q}\!\cdot\!\bbox{\nabla}_{1}\times\bbox{\nabla}_{2}]\,
    (\bbox{\sigma}_{1}+\bbox{\sigma}_{2})\!\cdot\!\bbox{\nabla}_{1}
    F(r_{1})G(r_{2}) = \frac{2}{r^{2}}F'G'\,{\bf L}\!\cdot\!{\bf S}.
\end{eqnarray*}

\noindent 3. For the nonadiabatic and pseudovector-vertex corrections:
\begin{eqnarray*}
 & & \lim_{{\bf r}_{1}\rightarrow{\bf r}_{2}}
   (\bbox{\nabla}_{1}\!\cdot\!\bbox{\nabla}_{2})^{3}F(r_{1})G(r_{2}) =
   \frac{6}{r^{2}}\left(F''-\frac{1}{r}F'\right)
                  \left(G''-\frac{1}{r}G'\right) + F'''G''',        \\
 & & \lim_{{\bf r}_{1}\rightarrow {\bf r}_{2}}
   (\bbox{\nabla}_{1}\!\cdot\!\bbox{\nabla}_{2})
   (\bbox{\sigma}_{1}\!\cdot\!\bbox{\nabla}_{1}\!\times\!\bbox{\nabla}_{2})
   (\bbox{\sigma}_{2}\!\cdot\!\bbox{\nabla}_{1}\!\times\!\bbox{\nabla}_{2})
        F(r_{1})G(r_{2})                                            \\
 & & \hspace{3.6cm}= -\frac{2}{3}\left[ \frac{1}{r^{2}}
        \left(\frac{1}{r}F'-F''+rF'''\right)
        \left(\frac{1}{r}G'-G''+rG'''\right) - F'''G'''\right]
   (\bbox{\sigma}_{1}\!\cdot\!\bbox{\sigma}_{2})                    \\
 & & \hspace{4.0cm}  +\frac{1}{3}\left[ \frac{1}{r^{2}}
        \left(\frac{2}{r}F'-2F''+\frac{r}{2}F'''\right)
        \left(\frac{2}{r}G'-2G''+\frac{r}{2}G'''\right)
        - \frac{1}{4}F'''G'''\right] S_{12},                        \\
 & & \lim_{{\bf r}_{1}\rightarrow{\bf r}_{2}}
    (\bbox{\nabla}_{1}\!\cdot\!\bbox{\nabla}_{2})
    (\bbox{\sigma}_{1}+\bbox{\sigma}_{2})\!\cdot\!
    (\bbox{\nabla}_{1}\times\bbox{\nabla}_{2})\,
    [{\bf q}\!\cdot\!(\bbox{\nabla}_{1}-\bbox{\nabla}_{2})]
    F(r_{1})G(r_{2}) =
\\ && \hspace{7.5cm}
    -\frac{4}{r^{2}}
       \left(F''-\frac{1}{r}F'\right)\left(G''-\frac{1}{r}G'\right)
      {\bf L}\!\cdot\!{\bf S},                                      \\
 & & \lim_{{\bf r}_{1}\rightarrow{\bf r}_{2}}
    (\bbox{\sigma}_{1}\!\cdot\!\bbox{\nabla}_{1})
    (\bbox{\sigma}_{2}\!\cdot\!\bbox{\nabla}_{2}) F(r_{1})G(r_{2}) =
    \frac{1}{3} F'G' \left[
     (\bbox{\sigma}_{1}\!\cdot\!\bbox{\sigma}_{2})+S_{12}\right],   \\
 & & \lim_{{\bf r}_{1}\rightarrow{\bf r}_{2}}
    (\bbox{\sigma}_{1}\!\cdot\!\bbox{\nabla}_{2})
    (\bbox{\sigma}_{2}\!\cdot\!\bbox{\nabla}_{1}) F(r_{1})G(r_{2}) =
    \frac{1}{3} F'G' \left[
     (\bbox{\sigma}_{1}\!\cdot\!\bbox{\sigma}_{2})+S_{12}\right],   \\
 & & \lim_{{\bf r}_{1}\rightarrow{\bf r}_{2}}
    (\bbox{\sigma}_{1}+\bbox{\sigma}_{2})\!\cdot\!{\bf q}\times
      (\bbox{\nabla}_{1}+\bbox{\nabla}_{2})\,
      (\bbox{\nabla}_{1}\!\cdot\!\bbox{\nabla}_{2}) F(r_{1})G(r_{2}) =
    -\frac{2}{r}\left[F''G'+F'G''\right]\, {\bf L}\!\cdot\!{\bf S}, \\
 & & \lim_{{\bf r}_{1}\rightarrow{\bf r}_{2}}
    (\bbox{\sigma}_{1}+\bbox{\sigma}_{2})\!\cdot\!
    (\bbox{\nabla}_{1}\times\bbox{\nabla}_{2})\, [{\bf q}\!\cdot\!
    (\bbox{\nabla}_{1}-\bbox{\nabla}_{2})] F(r_{1})G(r_{2}) =
    -\frac{4}{r^{2}}F'G'\, {\bf L}\!\cdot\!{\bf S}.
\end{eqnarray*}

\narrowtext
%\appendix
\section{\protect\\ COORDINATE-SPACE POTENTIALS}
\label{app:appB}
In this appendix we give the explicit expressions for the
coordinate-space two-meson-exchange potentials. We also
include the result for the two-pion-exchange potential from
Ref.~\cite{Rij91} for reasons of completeness.
In order to keep the expressions as compact as possible, we restrict
ourselves to the leading-order terms (no $1/M^{2}$ contributions),
and define $\odot$ and $\otimes$ operations as
\begin{eqnarray*}
  (F \odot G)_{\sigma}(r)   &=& \frac{1}{3}
                 \left(F''+\frac{2}{r}F'\right)G(r),                  \\
  (F \odot G)_{T}(r)        &=& \frac{1}{3}
                 \left(F''-\frac{1}{r}F'\right)G(r),                  \\
  (F \otimes G)_{C}(r)      &=&
                 \left[F''G''+\frac{2}{r^{2}}F'G'\right](r),          \\
  (F \otimes G)_{\sigma}(r) &=& \frac{2}{3} \left[\frac{1}{r^{2}}F'G'
               +\frac{1}{r}F''G'+\frac{1}{r}F'G''\right](r),          \\
  (F \otimes G)_{T}(r)      &=& \frac{1}{3} \left[\frac{2}{r^{2}}F'G'
               -\frac{1}{r}F''G'-\frac{1}{r}F'G''\right](r).
\end{eqnarray*}
We then find the following
[$F_{\alpha}(\lambda,r)$ is given by Eq.~(\ref{basicF})].

\noindent 1. Pseudoscalar exchanges:
\begin{eqnarray}
  V(\pi\pi) &=& -\left(\frac{f_{N\!N\pi}}{m_{\pi}}\right)^{4}\frac{2}{\pi}
            \int_{0}^{\infty}\!\frac{d\lambda}{\lambda^{2}}   \nonumber\\
  &\times&  \Biggl\{ 2(\bbox{\tau}_{1}\!\cdot\!\bbox{\tau}_{2})
         \biggl[\left(I_{2}^{\pi}\otimes I_{2}^{\pi}\right)_{C}
           -\left(F_{\pi}\otimes F_{\pi}\right)_{C}\biggr]    \nonumber\\
  &&  +3 \biggl[\left(I_{2}^{\pi}\otimes I_{2}^{\pi}\right)_{\sigma}
           -\left(F_{\pi}\otimes F_{\pi}\right)_{\sigma}\biggr]
           (\bbox{\sigma}_{1}\!\cdot\!\bbox{\sigma}_{2})      \nonumber\\
  &&  +3 \biggl[\left(I_{2}^{\pi}\otimes I_{2}^{\pi}\right)_{T}
      -\left(F_{\pi}\otimes F_{\pi}\right)_{T}\biggr] S_{12} \Biggr\}, \\
  V(\pi\eta) &=& -2\left(\frac{f_{N\!N\eta}}{m_{\pi}}\right)^{2}
                   \left(\frac{f_{N\!N\pi}}{m_{\pi}}\right)^{2}
             (\bbox{\tau}_{1}\!\cdot\!\bbox{\tau}_{2}) \frac{2}{\pi}
       \int_{0}^{\infty}\!\frac{d\lambda}{\lambda^{2}}        \nonumber\\
  &\times&  \Biggl\{\biggl[\left(I_{2}^{\pi}\otimes
                        I_{2}^{\eta}\right)_{\sigma}
           -\left(F_{\pi}\otimes F_{\eta}\right)_{\sigma}\biggr]
           (\bbox{\sigma}_{1}\!\cdot\!\bbox{\sigma}_{2})      \nonumber\\
  &&  + \biggl[\left(I_{2}^{\pi}\otimes I_{2}^{\eta}\right)_{T}
      -\left(F_{\pi}\otimes F_{\eta}\right)_{T}\biggr] S_{12}\Biggr\}, \\
  V(\pi\eta') &=& V(\pi\eta)\ \ {\rm with}\ \ \eta\rightarrow\eta'.
\end{eqnarray}
2. Scalar exchanges:
\begin{eqnarray}
  V(\pi a_{0}) &=& 4 g_{a_{0}}^{2}
                  \left(\frac{f_{N\!N\pi}}{m_{\pi}}\right)^{2}
             (\bbox{\tau}_{1}\!\cdot\!\bbox{\tau}_{2}) \frac{2}{\pi}
       \int_{0}^{\infty}\!\frac{d\lambda}{\lambda^{2}}       \nonumber\\
  &\times&   \Biggl\{\biggl[\left(I_{2}^{\pi}\odot
                  I_{2}^{a_{0}}\right)_{\sigma}
       -\left(F_{\pi}\odot F_{a_{0}}\right)_{\sigma}\biggr]
       (\bbox{\sigma}_{1}\!\cdot\!\bbox{\sigma}_{2})         \nonumber\\
  &&  + \biggl[\left(I_{2}^{\pi}\odot I_{2}^{a_{0}}\right)_{T}
      -\left(F_{\pi}\odot F_{a_{0}}\right)_{T}\biggr]S_{12} \Biggr\}, \\
  V(\pi\varepsilon) &=& V(\pi f_{0}) = 0.
\end{eqnarray}
3. Pomeron exchange:
\begin{eqnarray}
  V(\pi P) &=& 0.
\end{eqnarray}
4. Vector exchanges:
\begin{eqnarray}
  V(\pi\rho) &=& -4 g_{\rho}^{2}
                  \left(\frac{f_{N\!N\pi}}{m_{\pi}}\right)^{2}
             (\bbox{\tau}_{1}\!\cdot\!\bbox{\tau}_{2}) \frac{2}{\pi}
       \int_{0}^{\infty}\!\frac{d\lambda}{\lambda^{2}}       \nonumber\\
  &\times&   \Biggl\{\biggl[\left(I_{2}^{\pi}\odot
                  I_{2}^{\rho}\right)_{\sigma}
       -\left(F_{\pi}\odot F_{\rho}\right)_{\sigma}\biggr]
       (\bbox{\sigma}_{1}\!\cdot\!\bbox{\sigma}_{2})         \nonumber\\
  &&  + \biggl[\left(I_{2}^{\pi}\odot I_{2}^{\rho}\right)_{T}
      -\left(F_{\pi}\odot F_{\rho}\right)_{T}\biggr] S_{12} \Biggr\}, \\
  V(\pi\omega) &=& V(\pi\phi) = 0.   \label{B8}
\end{eqnarray}

\narrowtext

\narrowtext
\begin{table}
\caption{Adiabatic approximation of the energy denominators
         $D_{i}(\omega_{1},\omega_{2})$ for the planar and
         crossed diagrams.}
\begin{tabular}{lr}
 planar ($//$)    &  $+{\displaystyle
            \frac{1}{2\omega_{1}^{2}\omega_{2}^{2}}\
             \left[\frac{1}{\omega_{1}}+\frac{1}{\omega_{2}}
                  -\frac{1}{(\omega_{1}+\omega_{2})} \right]} $ \\ [0.5cm]
 crossed ($X$)    &  $-{\displaystyle
            \frac{1}{2\omega_{1}^{2}\omega_{2}^{2}}\
             \left[\frac{1}{\omega_{1}}+\frac{1}{\omega_{2}}
                  -\frac{1}{(\omega_{1}+\omega_{2})} \right]} $
\end{tabular}
\label{table1}
\end{table}

\widetext
\begin{table}
\caption{The momentum operators $O^{(i)}({\bf k}_{1},{\bf k}_{2})$
         for the planar (//) and crossed ($X$) $\pi$-meson-exchange
         diagrams. The subscripts $PS$ and $S$ refer to the
         $\pi$-pseudoscalar and $\pi$-scalar operators, respectively.
         The subscripts $e$ and $m$ refer to the electric
         and magnetic parts of the $\pi$-vector operators, and
         $\kappa=f_{V}/g_{V}$. The $\mp$ sign refers to pseudovector
         or pseudoscalar coupling for the pion, respectively.}
\begin{tabular}{cl}
 Type & \multicolumn{1}{c}{$O^{(i)}({\bf k}_{1},{\bf k}_{2})$} \\
 \tableline \\
 $O^{(//)}_{PS} $ & $2({\bf k}_{1}\!\cdot\!{\bf k}_{2})^{2} - 2
     \bbox{\sigma}_{1}\!\cdot\!({\bf k}_{1}\!\times\!{\bf k}_{2})
     \bbox{\sigma}_{2}\!\cdot\!({\bf k}_{1}\!\times\!{\bf k}_{2})$ \\ [0.2cm]
 $O^{(X)}_{PS} $  & $2({\bf k}_{1}\!\cdot\!{\bf k}_{2})^{2} + 2
     \bbox{\sigma}_{1}\!\cdot\!({\bf k}_{1}\!\times\!{\bf k}_{2})
     \bbox{\sigma}_{2}\!\cdot\!({\bf k}_{1}\!\times\!{\bf k}_{2})$ \\
    & \\
 \tableline  \\
 $O^{(//)}_{S} $ & $ {\displaystyle\left[
      2+\frac{{\bf k}_{2}^{2}+{\bf k}_{1}\!\cdot\!{\bf k}_{2}}{2M^{2}}
      -\frac{{\bf q}^{2}+{\textstyle\frac{1}{4}}{\bf k}^{2}}{M^{2}}
      \right] \bbox{\sigma}_{1}\!\cdot\!{\bf k}_{1}
              \bbox{\sigma}_{2}\!\cdot\!{\bf k}_{1}
}$ \\ [0.4cm] & $ {\displaystyle\hspace{1cm}
        - \frac{1}{4M^{2}} \biggl[
      ({\bf k}_{1}^{2}\mp{\bf k}_{1}\!\cdot\!{\bf k}_{2})
                     (\bbox{\sigma}_{1}\!\cdot\!{\bf k}_{1}
                      \bbox{\sigma}_{2}\!\cdot\!{\bf k}_{2}
                     +\bbox{\sigma}_{1}\!\cdot\!{\bf k}_{2}
                      \bbox{\sigma}_{2}\!\cdot\!{\bf k}_{1})
      + 2i{\bf q}\!\cdot\!({\bf k}_{1}\!\times\!{\bf k}_{2})
      (\bbox{\sigma}_{1}+\bbox{\sigma}_{2})\!\cdot\!{\bf k}_{1} \biggr]}$
                                                             \\ [0.4cm]
 $O^{(X)}_{S} $ & $ {\displaystyle\left[
      2+\frac{{\bf k}_{2}^{2}+{\bf k}_{1}\!\cdot\!{\bf k}_{2}}{2M^{2}}
      -\frac{{\bf q}^{2}+{\textstyle\frac{1}{4}}{\bf k}^{2}}{M^{2}}
      \right] \bbox{\sigma}_{1}\!\cdot\!{\bf k}_{1}
              \bbox{\sigma}_{2}\!\cdot\!{\bf k}_{1}
}$ \\ [0.4cm] & $ {\displaystyle\hspace{1cm}
        - \frac{1}{4M^{2}} \biggl[
      ({\bf k}_{1}^{2}\mp{\bf k}_{1}\!\cdot\!{\bf k}_{2})
                     (\bbox{\sigma}_{1}\!\cdot\!{\bf k}_{1}
                      \bbox{\sigma}_{2}\!\cdot\!{\bf k}_{2}
                     +\bbox{\sigma}_{1}\!\cdot\!{\bf k}_{2}
                      \bbox{\sigma}_{2}\!\cdot\!{\bf k}_{1})
      + 2i{\bf q}\!\cdot\!({\bf k}_{1}\!\times\!{\bf k}_{2})
      (\bbox{\sigma}_{1}+\bbox{\sigma}_{2})\!\cdot\!{\bf k}_{1}\biggr]}$\\
  & \\
 \tableline  \\
 $O^{(//)}_{e,e}$ & ${\displaystyle\left[ -2
      -3\frac{{\bf q}^{2}+{\textstyle\frac{1}{4}}{\bf k}^{2}}{M^{2}}
     +\frac{{\bf k}_{1}^{2}+4{\bf k}_{1}\!\cdot\!{\bf k}_{2}
           +{\bf k}_{2}^{2}}{2M^{2}}-\frac{{\bf k}_{1}^{2}}{2M^{2}}
     \right] \bbox{\sigma}_{1}\!\cdot\!{\bf k}_{1}
             \bbox{\sigma}_{2}\!\cdot\!{\bf k}_{1}
     \mp \frac{{\bf k}_{1}\!\cdot\!{\bf k}_{2}}{4M^{2}}
              (\bbox{\sigma}_{1}\!\cdot\!{\bf k}_{1}
               \bbox{\sigma}_{2}\!\cdot\!{\bf k}_{2}
              +\bbox{\sigma}_{1}\!\cdot\!{\bf k}_{2}
               \bbox{\sigma}_{2}\!\cdot\!{\bf k}_{1}) } $  \\ [0.4cm]
 $O^{(X)}_{e,e}$ & ${\displaystyle\left[ -2
      -3\frac{{\bf q}^{2}+{\textstyle\frac{1}{4}}{\bf k}^{2}}{M^{2}}
     +\frac{{\bf k}_{1}^{2}+4{\bf k}_{1}\!\cdot\!{\bf k}_{2}
           +{\bf k}_{2}^{2}}{2M^{2}}+\frac{{\bf k}_{1}^{2}}{2M^{2}}
     \right] \bbox{\sigma}_{1}\!\cdot\!{\bf k}_{1}
             \bbox{\sigma}_{2}\!\cdot\!{\bf k}_{1}
     \mp \frac{{\bf k}_{1}\!\cdot\!{\bf k}_{2}}{4M^{2}}
              (\bbox{\sigma}_{1}\!\cdot\!{\bf k}_{1}
               \bbox{\sigma}_{2}\!\cdot\!{\bf k}_{2}
              +\bbox{\sigma}_{1}\!\cdot\!{\bf k}_{2}
               \bbox{\sigma}_{2}\!\cdot\!{\bf k}_{1}) } $  \\
    & \\
 $O^{(//)}_{e,m}$ & $(1+\kappa) \left[
     -2i{\bf q}\!\cdot\!({\bf k}_{1}\!\times\!{\bf k}_{2})
       (\bbox{\sigma}_{1}+\bbox{\sigma}_{2})\!\cdot\!{\bf k}_{1}
     +2{\bf k}_{1}\!\cdot\!{\bf k}_{2}
        \bbox{\sigma}_{1}\!\cdot\!{\bf k}_{1}
        \bbox{\sigma}_{2}\!\cdot\!{\bf k}_{1} - {\bf k}_{1}^{2}
       (\bbox{\sigma}_{1}\!\cdot\!{\bf k}_{1}
        \bbox{\sigma}_{2}\!\cdot\!{\bf k}_{2}
       +\bbox{\sigma}_{1}\!\cdot\!{\bf k}_{2}
        \bbox{\sigma}_{2}\!\cdot\!{\bf k}_{1}) \right] $    \\ [0.2cm]
 $O^{(X)}_{e,m}$ & $(1+\kappa) \left[
     -2i{\bf q}\!\cdot\!({\bf k}_{1}\!\times\!{\bf k}_{2})
       (\bbox{\sigma}_{1}+\bbox{\sigma}_{2})\!\cdot\!{\bf k}_{1}
     -2{\bf k}_{1}\!\cdot\!{\bf k}_{2}
        \bbox{\sigma}_{1}\!\cdot\!{\bf k}_{1}
        \bbox{\sigma}_{2}\!\cdot\!{\bf k}_{1} + {\bf k}_{1}^{2}
       (\bbox{\sigma}_{1}\!\cdot\!{\bf k}_{1}
        \bbox{\sigma}_{2}\!\cdot\!{\bf k}_{2}
       +\bbox{\sigma}_{1}\!\cdot\!{\bf k}_{2}
        \bbox{\sigma}_{2}\!\cdot\!{\bf k}_{1}) \right] $ \\
    & \\
 $O^{(//)}_{e,2m}$ & $(1+2\kappa) \left[
     -i{\bf q}\!\cdot\!({\bf k}_{1}\!\times\!{\bf k}_{2})
      (\bbox{\sigma}_{1}+\bbox{\sigma}_{2})\!\cdot\!{\bf k}_{1}
    +({\bf k}_{2}^{2}+{\bf k}_{1}\!\cdot\!{\bf k}_{2})
       \bbox{\sigma}_{1}\!\cdot\!{\bf k}_{1}
       \bbox{\sigma}_{2}\!\cdot\!{\bf k}_{1}
\right. $ \\ [0.2cm] & $ \hspace{1.7cm} \left.
     -{\textstyle\frac{1}{2}}{\bf k}_{1}^{2}
      (\bbox{\sigma}_{1}\!\cdot\!{\bf k}_{1}
       \bbox{\sigma}_{2}\!\cdot\!{\bf k}_{2}
      +\bbox{\sigma}_{1}\!\cdot\!{\bf k}_{2}
       \bbox{\sigma}_{2}\!\cdot\!{\bf k}_{1}) \right] $    \\ [0.2cm]
 $O^{(X)}_{e,2m}$ & $(1+2\kappa) \left[
     -i{\bf q}\!\cdot\!({\bf k}_{1}\!\times\!{\bf k}_{2})
       (\bbox{\sigma}_{1}+\bbox{\sigma}_{2})\!\cdot\!{\bf k}_{1}
     +({\bf k}_{2}^{2}+{\bf k}_{1}\!\cdot\!{\bf k}_{2})
        \bbox{\sigma}_{1}\!\cdot\!{\bf k}_{1}
        \bbox{\sigma}_{2}\!\cdot\!{\bf k}_{1}
\right. $ \\ [0.2cm] & $ \hspace{1.7cm} \left.
     -{\textstyle\frac{1}{2}}{\bf k}_{1}^{2}
      (\bbox{\sigma}_{1}\!\cdot\!{\bf k}_{1}
       \bbox{\sigma}_{2}\!\cdot\!{\bf k}_{2}
      +\bbox{\sigma}_{1}\!\cdot\!{\bf k}_{2}
       \bbox{\sigma}_{2}\!\cdot\!{\bf k}_{1}) \right] $      \\
    & \\
 $O^{(//)}_{m,m}$ & $ (1+\kappa)^{2} \left[
      {\bf k}_{1}^{2}{\bf k}_{2}^{2}
     -({\bf k}_{1}\!\cdot\!{\bf k}_{2})^{2}
    + {\bf k}_{2}^{2}\bbox{\sigma}_{1}\!\cdot\!{\bf k}_{1}
                     \bbox{\sigma}_{2}\!\cdot\!{\bf k}_{1}
    - {\bf k}_{1}^{2}{\bf k}_{2}^{2}
                      \bbox{\sigma}_{1}\!\cdot\!\bbox{\sigma}_{2}
    + \bbox{\sigma}_{1}\!\cdot\!({\bf k}_{1}\!\times\!{\bf k}_{2})
      \bbox{\sigma}_{2}\!\cdot\!({\bf k}_{1}\!\times\!{\bf k}_{2})
      \right]$                                              \\ [0.2cm]
 $O^{(X)}_{m,m}$ & $ (1+\kappa)^{2} \left[
      {\bf k}_{1}^{2}{\bf k}_{2}^{2}
     -({\bf k}_{1}\!\cdot\!{\bf k}_{2})^{2}
     - {\bf k}_{2}^{2}\bbox{\sigma}_{1}\!\cdot\!{\bf k}_{1}
                      \bbox{\sigma}_{2}\!\cdot\!{\bf k}_{1}
    + {\bf k}_{1}^{2}{\bf k}_{2}^{2}
                      \bbox{\sigma}_{1}\!\cdot\!\bbox{\sigma}_{2}
    - \bbox{\sigma}_{1}\!\cdot\!({\bf k}_{1}\!\times\!{\bf k}_{2})
      \bbox{\sigma}_{2}\!\cdot\!({\bf k}_{1}\!\times\!{\bf k}_{2})\right]$
\end{tabular}
\label{table2}
\end{table}

\narrowtext
\begin{table}
\caption{Energy denominators $D_{i}^{(1)}(\omega_{1},\omega_{2})$ for
         the nonadiabatic corrections to the planar and crossed
         diagrams.}
\begin{tabular}{lr}
 planar   ($//$)  &  $+{\displaystyle
             \frac{1}{2\omega_{1}^{2}\omega_{2}^{2}}\ \left[
    \frac{1}{\omega_{1}^{2}}+\frac{1}{\omega_{2}^{2}}\right] }$ \\ [0.5cm]
 crossed ($X$)    &  $-{\displaystyle
             \frac{1}{\omega_{1}^{2}\omega_{2}^{2}}\  \left[
    \frac{1}{\omega_{1}^{2}}+\frac{1}{\omega_{2}^{2}}\right] }$
\end{tabular}
\label{table3}
\end{table}

\narrowtext
\begin{table}
\caption{Meson parameters employed in the potentials shown in
         Figs.~\protect\ref{pap1fig3} to \protect\ref{pap1fig8}.
         Coupling constants are at ${\bf k}^{2}=0$.}
\begin{tabular}{cdrrrd}
 meson & mass (MeV) & $g/\sqrt{4\pi}$ & $f/\sqrt{4\pi}$
       & $\kappa=f/g$ & $\Lambda$ (MeV) \\
\tableline
 $\pi$         &  138.04 &         &   0.269  &       &  853.9  \\
 $\eta$        &  547.45 &         &   0.124  &       &  950.0  \\
 $\eta'$       &  957.75 &         &   0.124  &       &  950.0  \\
 $\rho$        &  768.10 &  0.622  &   2.306  &  3.71 &  886.2  \\
 $\omega$      &  781.95 &  3.336  & --0.206  &--0.06 & 1013.1  \\
 $\phi$        & 1019.41 &--1.202  &   0.078  &--0.07 & 1013.1  \\
 $a_{0}$       &  982.70 &  1.847  &          &       &  679.4  \\
 $\varepsilon$ &  760.00 &  0.582  &          &       &  759.0  \\
 $f_{0}$       &  974.10 &--2.372  &          &       &  759.0  \\
 Pomeron       &  309.10 &  1.884  &          &       &
\end{tabular}
\label{table4}
\end{table}

\narrowtext
\begin{figure}
\caption{BW two-meson-exchange graphs: (a) planar and (b)--(d) crossed
         box. The dotted line with momentum ${\bf k}_{1}$ refers to the
         pion and the dashed line with momentum ${\bf k}_{2}$ refers
         to one of the other (vector, scalar, or pseudoscalar) mesons.
         To these we have to add the ``mirror'' graphs, and the
         graphs where we interchange the two meson lines.}
\label{pap1fig1}
\end{figure}

\begin{figure}
\caption{Planar-box TMO two-meson-exchange graphs.
         Same notation as in Fig.~\protect\ref{pap1fig1}.
         To these we have to add the ``mirror'' graphs, and the
         graphs where we interchange the two meson lines.}
\label{pap1fig2}
\end{figure}

\begin{figure}
\caption{Spin-orbit contributions of the pion-meson potentials
         for both $I=0$ and $I=1$ in comparison with the
         one-boson-exchange contribution.}
\label{pap1fig8}
\end{figure}

\end{document}